\documentclass[prb,aps,showpacs,twocolumn]{revtex4}
\usepackage{amsmath}
\usepackage{graphicx}
\usepackage{color}
\begin{document}
\title{Phase-sensitive transport at a
  normal metal-superconductor interface close to a Josephson junction}
\date{\today}

\author{David Gosselin, Gaston Hornecker, R\'egis M\'elin, Denis Feinberg}
\email{denis.feinberg@neel.cnrs.fr}
\affiliation{Universit\'e Grenoble-Alpes, Institut NEEL,
 F-38042 Grenoble Cedex 9, France}
\affiliation{CNRS, Institut NEEL,
 F-38042 Grenoble Cedex 9, France}

\begin{abstract}
Phase- and voltage bias-sensitive quasiparticle transport at a double
$NIS_1IS_2$ interface is considered. The barriers $I$ range 
from tunnel to transparent, and the intermediate region $S_1$ has a width comparable to the
superconducting coherence length. A phase difference $\varphi$ is applied to
the Josephson junction $S_1IS_2$. The normal and Andreev reflections at the
$NIS_1$ interface become $\varphi$-sensitive, and transport is governed by
interferences within the narrow $S_1$ region, both in the normal and anomalous
channels. The subgap conductance is separately (energy $E$)- and (phase
$\varphi$)- symmetric. Above the superconducting gap, the conductance is
in general not symmetric even if $(E,\varphi)$
is changed in $(-E,-\varphi)$, but the symmetry is restored by averaging Fermi oscillations. 
The Tomasch oscillations are amplified
by the phase difference. The subgap conductance exhibits a resonant structure
at the energy of the Andreev bound states (ABS) of the $S_1IS_2$ junction,
providing a side-spectroscopy of such states. Depending on the relative
transparencies of the junctions, the resonance can increase or reduce the conductance, 
and it can even vanish for $\varphi=\pi$, featuring total
reflection of quasiparticles at $NS_1$ by the ABS at $S_1S_2$.
\end{abstract}

\maketitle

\section{Introduction}
Transport in hybrid set-ups involving {interfaces} between superconductors
($S$) and normal metals ($N$) is governed by Andreev reflection, where an
incoming electron with energy $\mu+E$ is transformed into a hole with opposite energy 
$\mu-E$ in the
metal ($\mu$ is the chemical potential)\cite{Andreev,SaintJames}, and a Cooper pair enters the superconducting
condensate. Conversely, an incoming hole may be reflected as an electron
  while a pair is taken from $S$.  Andreev scattering dominates subgap
transport and it plays also a role at energies of order of a few times
the superconducting gap. Using the Bogoliubov-De Gennes equations and writing
the scattering equations for the electron and hole wavefunctions, De Gennes
and Saint-James found subgap bound states in a thin metallic layer in
contact with a superconductor \cite{SaintJames}. Rowell and McMillan
\cite{RowellMcMillan} showed that conductance oscillations occur as well above
the gap. Tomasch\cite{Tomasch} discovered oscillations in the conductance
above the gap in a $NSIN$ structure, which were explained by McMillan and
Anderson \cite{McMillanAnderson} as an interference effect due to the
wavevector mismatch between the electron and hole-like
quasiparticle branches propagating in a narrow $S$ layer of thickness $L$ of
the order of the superconducting coherence length $\xi$.

Using the scattering approach, Blonder, Tinkham and Klapwijk
\cite{BTK} were able to bridge the gap between a
Giaever tunneling barrier ($NIS$) and a perfectly
transparent $NS$ interface, where the conductance is doubled below the gap,
with respect to the normal case. The scattering approach had
been used previously for a double $SNS$ interface by Kulik
\cite{Kulik} in the transparent case, then in many
subsequent works, to obtain a complete description of a
clean $SINIS$ Josephson junction. It is characterized by the formation of
Andreev bound states (ABS), as resonant states formed by
multiple electron-electron and electron-hole scattering at
each $SIN$ interface. The phase dispersion of the ABS is
responsible for the Josephson current flowing through the junction. Recently,
a tunnel spectroscopy of the ABS was performed by attaching
a third contact to the normal bridge of a long junction \cite{Pillet}. A
microwave spectroscopy of the ABS was recently obtained in an
atomic point contact \cite{Saclay,Bretheau} and a diffusive metallic junction
\cite{Bouchiat}.

These effects have been probed
experimentally in two-terminal transport geometries. In addition, tunnel
spectroscopy of the ABS involves a third reservoir, weakly coupled directly to the junction
\cite{Pillet,Giazotto}. Coupling a $SNS$ Josephson junction to a normal wire has also
been proposed and achieved \cite{SSNN}. Recently, new three-terminal hybrid
configurations {exploiting} the mesoscopic size of Cooper
pairs have been explored. For instance, consider a $NISIN$ geometry where two
normal leads (or quantum dots) are connected by a narrow superconducting
region, of size comparable to the superconducting coherence length
$\xi$. Then, a mechanism denoted as crossed (or nonlocal) Andreev reflection
allows a hole {incoming} on one side to be transmitted as an
electron on the other, effectively splitting a Cooper pair from the
superconductor, into a pair of correlated quasiparticles \cite{CAR}. Those carry opposite
energies and spins (for an s-wave superconductor), and this mechanism has been
proposed as a source of entangled fermions in the solid state \cite{Recher}. A
three-terminal all-superconducting $SISIS$ set-up, called as
  a Josephson bijunction, has been more recently considered
\cite{Cuevas2007,Houzet2010,Freyn2011,Jonckheere2013,Coupiac2011,Pfeffer2013}. It
was shown that the two independent phase degree of freedom (or voltages) 
leads to novel coherent multipair
dc channels, which coexist with dissipative quasiparticle transport for some
combinations of applied voltages.

In what follows, we investigate a hybrid structure where Andreev reflection interferes with Josephson transport at a neighbouring junction. 
It is controlled by one voltage and one phase, as {\it independent} variables. 
More specifically, we explore the properties of a hybrid bijunction
$NIS_1IS_2$, made of a $NIS_1$ interface, in close proximity with
a Josephson junction $S_1IS_2$. The transparencies of the two interfaces {are
  arbitrary. The Josephson junction is biased with a phase difference
$\varphi$. A possible experimental device is sketched on Fig.\ref{scheme}. Alternatively, the phase
difference can also be imposed by an
applied current, in a three-terminal geometry, but accessing only the range $\varphi=[-\pi/2, \pi/2]$. 

To modelize such a structure, we consider a one-dimensional scattering model with two interfaces in series. 
As a consequence, the currents at the interfaces $NIS_1$ et 
$S_1IS_2$ are not necessarily equal. The quasiparticle current in N is converted into a Cooper pair current flowing, 
partly through the junction, the other part in the upper branch of the loop. In spite of the presence of
interferences in this set-up, the geometry is very different from the
usual Andreev interferometer containing two $NS$ interfaces in parallel
\cite{Andreevinterferometer}. The conductance through the $NIS_1$ interface is
calculated within the scattering approach.  The
  phase difference between $S_1$ and $S_2$, as a new control variable, brings
  quantitative and qualitative changes as compared to previous calculations
  including scattering within a single superconductor
  \cite{Bagwell}. Resonant tunneling in a $NSNSN$ double barrier geometry was
also investigated in Ref. \onlinecite{Morpurgo}, focusing on the case of
ideal $NS$ interfaces.  Conversely, we consider here an asymmetric
$NS_1S_2$ structure where, in addition, the transparencies of the
$NS_1$ and $S_1S_2$ junctions are
arbitrary.

 In this work, we calculate the conductance $G(E,\varphi)$ as a function of
  the voltage energy $E=eV$ and the phase $\varphi$. This conductance is
  the derivative of the current through the $NIS_1$ interface with respect to
  the voltage $V$ on $N$, the two superconducting regions $S_1$ and $S_2$ being
  grounded but phase-biased.  A first result is that the
  energy and phase symmetries of the Andreev reflection probability are
  broken above the gap. Second, the Tomasch conductance oscillations become phase-sensitive
and they are amplified. Third, the subgap conductance displays a resonant
behavior close to the ABS state energies, yielding a novel ABS spectroscopy
tool. Changing the interface parameters, this resonance crosses-over between a conductance maximum, featuring
``transmission'' tunneling spectroscopy, and a
conductance minimum, featuring ``reflection''
spectroscopy. The first situation is encountered when the $NIS_1$ barrier is a
tunnel barrier or at least less transparent than the $S_1IS_2$ one, while  the
  second case, less conventional, corresponds to the converse where the
$NIS_1$ barrier is more transparent than the $S_1IS_2$ one. The perfect
cancellation of the conductance at $\varphi = \pi$ at the ABS energy is a
striking property, due to the suppression of the Andreev reflection by
interference between the two interfaces.

Sec. II presents the model and an analytical solution for
the simple limiting case of perfectly transparent
interfaces. Sec. III focuses on the conductance above the
gap. Sec. IV details the subgap
conductance. Sec. V provides a general discussion.

\begin{figure}
\centering
\includegraphics[width=8cm]{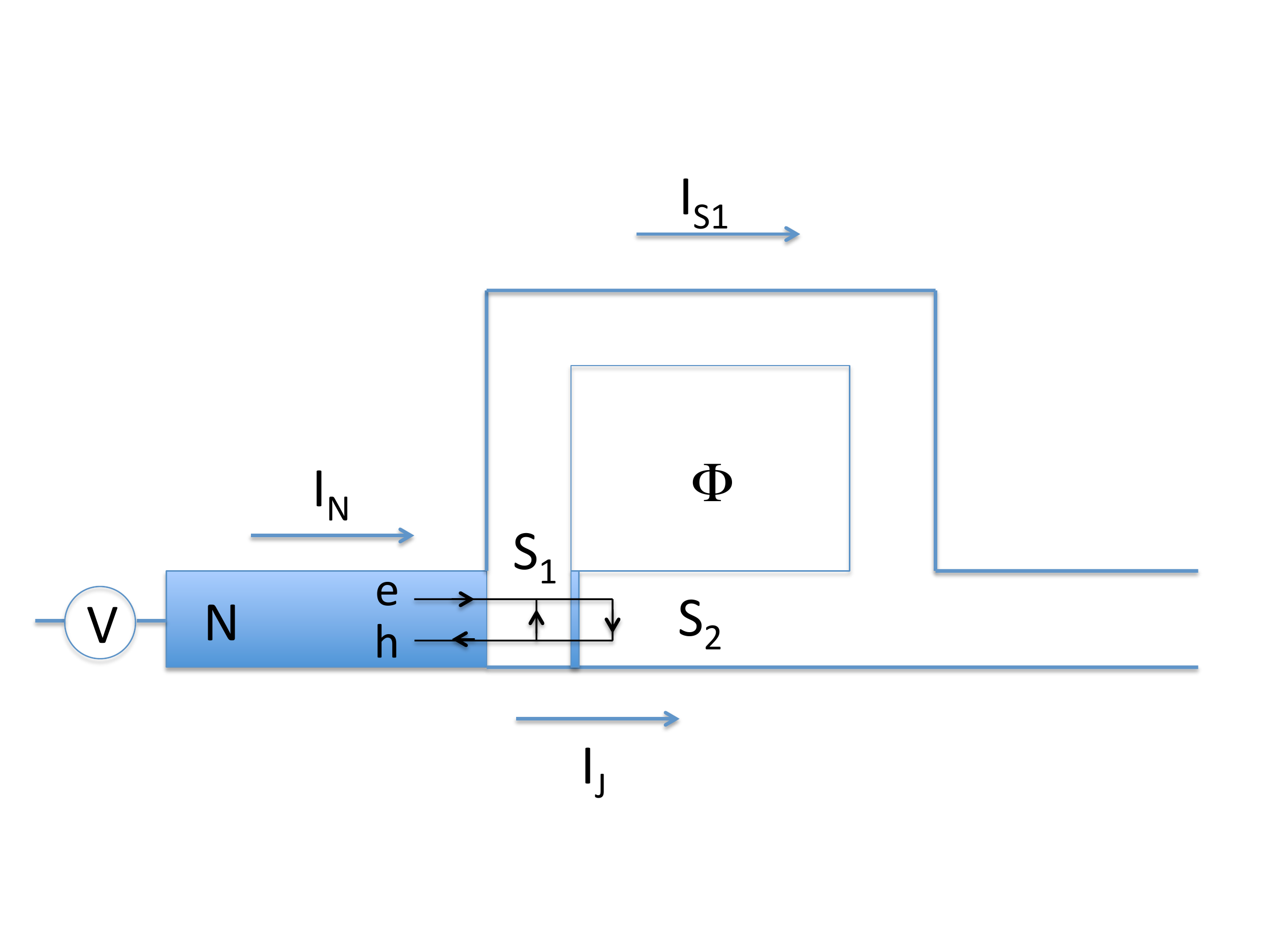}
\caption{\label{scheme}Schematics of an hybrid $NS$ loop set-up. A superconducting loop at zero voltage, cut by a Josephson junction, is connected on the left to a normal metal lead, at a voltage $V$, at a distance from the junction comparable to the coherence length (in practice the loop thickness is larger than pictured). The flux $\Phi$ imposes a phase difference $\varphi$ at the junction. The total current $I_N$ splits into a Josephson current $I_J$ and a current $I_{S1}$ flowing in the upper branch of the loop. The scattering model is one-dimensional and involves a double $NS_1S_2$ interface, where $S_1$ and $S_2$ are the parts of the (same) superconductor, set at phases $0$ and $\varphi$ by convention. The incident electron and scattered hole waves are represented, together with Andreev reflection processes within $S_1$, and those within $S_2$, close to the junction, that generates the Andreev bound states.}
\end{figure}

\section{The model}

\subsection{Matching equations}
A double $NIS_1IS_2$ interface is considered. For sake of simplicity, we assume that the Fermi energy 
and velocity are the same in all materials, and that the superconductors $S_1$ and $S_2$ have the same gaps 
(this corresponds to the scheme of Fig.\ref{scheme} where the superconducting loop is made of the same material). The barrier
  transparencies are defined as $Z_{1,2}=H_1/\hbar v_F$ where
$H_{1,2}$ are the amplitudes of delta-function barriers at the interfaces
\cite{BTK}. The phases are $\varphi_1=0$
  and $\varphi_2=\varphi$. This one-dimensional model can be extended to more realistic interfaces, as
  discussed at the end of the paper. In the case of $N$ and $S$ having different electronic parameters, this is known 
to quantitatively modify the scattering equations, making the $NS_1$ barrier less transparent. For instance, for a transparent interface, 
a wavevector mismatch $k_F/k_S \neq 1$ is exactly equivalent to an effective barrier $Z_{eff}$. 
The main conclusions of the work will thus not be qualitatively
modified, and the present calculation can be easily extended to take into account a parameter mismatch.  Following BTK \cite{BTK},
a right-moving electron wavefunction in $N$ is incoming onto
  the interface. The quasiparticle wavefunctions in $N$, $S_1$ and
$S_2$ can then be written as

\begin{equation} \label{eq_psi_e}
\begin{split}
\Psi_N^{e} \left(x\right) & = \left( \begin{matrix} 1 \\ 0 \end{matrix}
\right) \left( e^{i q^{+} x} + b\, e^{-iq^{+}x} \right) +
\left( \begin{matrix} 0 \\ 1 \end{matrix} \right) a\, e^{iq^{-}x}
\\ \Psi_{S_1}^{e} \left(x\right) & = \left( \begin{matrix} u_0
  \\ v_0 \end{matrix} \right) \left( \alpha_n\, e^{i k^{+} x} + \beta_n\,
e^{-ik^{+}x} \right)\\ &+ \left( \begin{matrix} v_0 \\ u_0 \end{matrix}
\right) \left( \alpha_a\, e^{i k^{-} x} + \beta_a\, e^{-ik^{-}x} \right)
\\ \Psi_{S_2}^{e} \left(x\right) & = \left( \begin{matrix} u_0 e^{i \varphi}
  \\ v_0 \end{matrix} \right) c\, e^{i k^{+} x} + \left( \begin{matrix} v_0
  e^{i \varphi} \\ u_0 \end{matrix} \right)d\, e^{-ik^{-}x},
\end{split}
\end{equation}

\noindent
where $a$, $b$, $\alpha_n$, $\beta_n$, $\alpha_a$, $\beta_a$, $c$, $d$ are
amplitude probabilities: $a$ (resp. $b$) for Andreev (resp. normal) reflection
to the left in $N$, $\alpha_n$ (resp. $\beta_a$) for electron-like
(resp. hole-like) right-moving waves in $S_1$, $\beta_n$ (resp. $\alpha_a$)
for electron-like (resp. hole-like) left-moving waves in $S_1$, $c$
(resp. $d$) for electron-like (resp. hole-like) right-moving waves in $S_2$
(see Fig. \ref{incident_e}). Here

\begin{equation} \label{eq_vect_onde}
\begin{split}
\hbar q^{\pm} & = \sqrt{2m (\mu \pm E)}, \;\;\; \hbar k^{\pm} = \sqrt{2m} [\mu \pm \sqrt{E^2 - \Delta^2}]^{1/2},\\
u_0 ^2 & = \dfrac{1}{2} \left( 1 + \dfrac{\sqrt{E^2 - \Delta^2}}{E} \right) = 1 - v_0 ^2,
\end{split}
\end{equation}

\noindent
with the {relation} $u_0(-E)=v_0(E)$ and
$v_0(-E)=-u_0(E)$. 

\begin{figure}
\centering
\includegraphics[width=6cm]{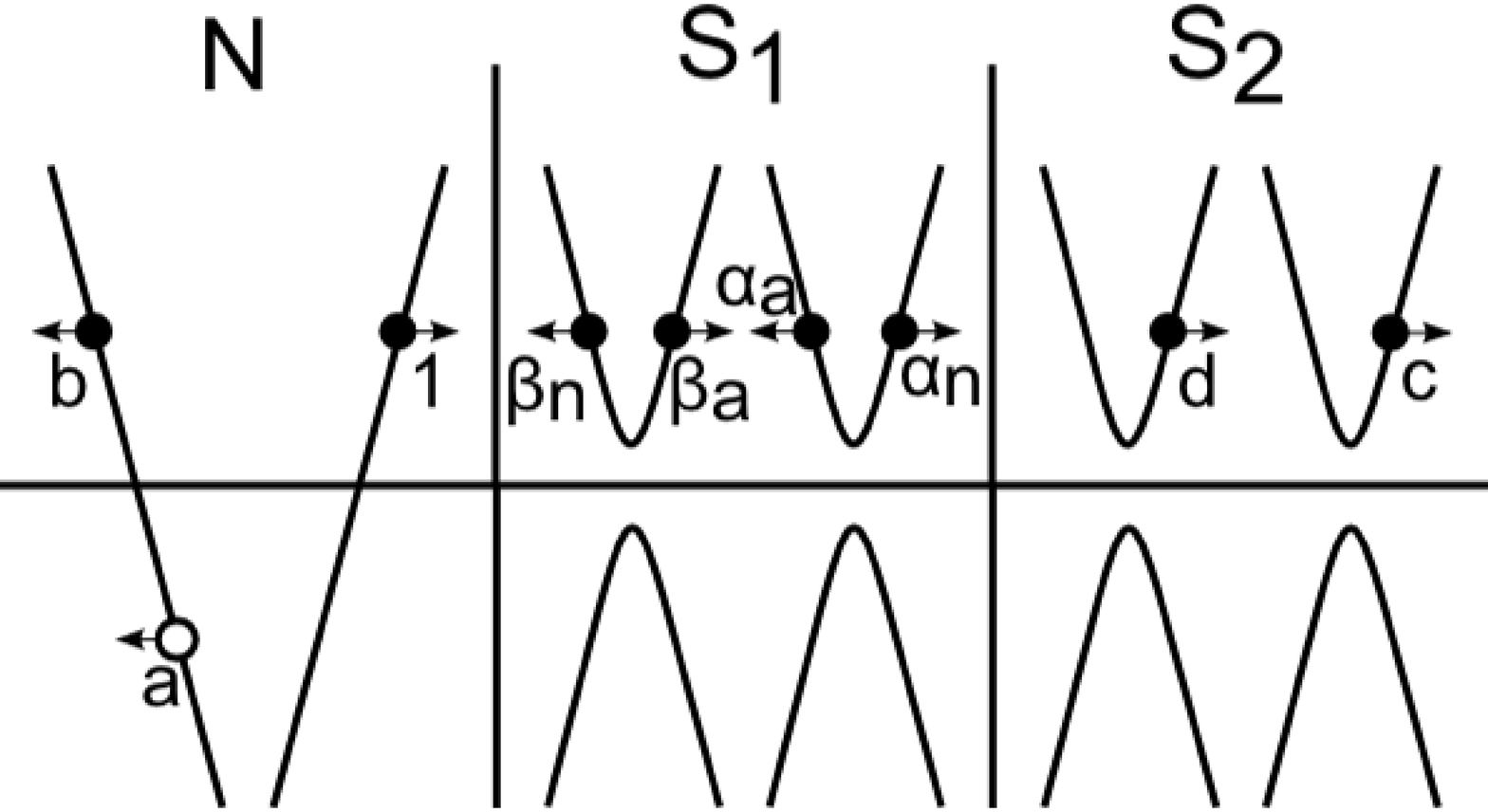}
\caption{\label{incident_e}Semiconductor diagram for an incident electron at the double {$NIS_1IS_2$} interface.}
\end{figure}

As shown below, generically, here for a nonzero phase
difference, the conductance is {\it not symmetric} in energy, e.g.
the standard BTK relation $A(E) = A(-E)$ for a single
  interface does not hold any longer in the considered set-up (see
  Table~\ref{table}). It is therefore
convenient to use Eq.~(\ref{eq_psi_e}) for $E>0$, and to use instead for $E
< 0$ similar equations for an incoming hole with amplitudes $\bar a, \bar b$ etc...
(see Fig. \ref{incident_h}):
\begin{figure}
\centering
\includegraphics[width=6cm]{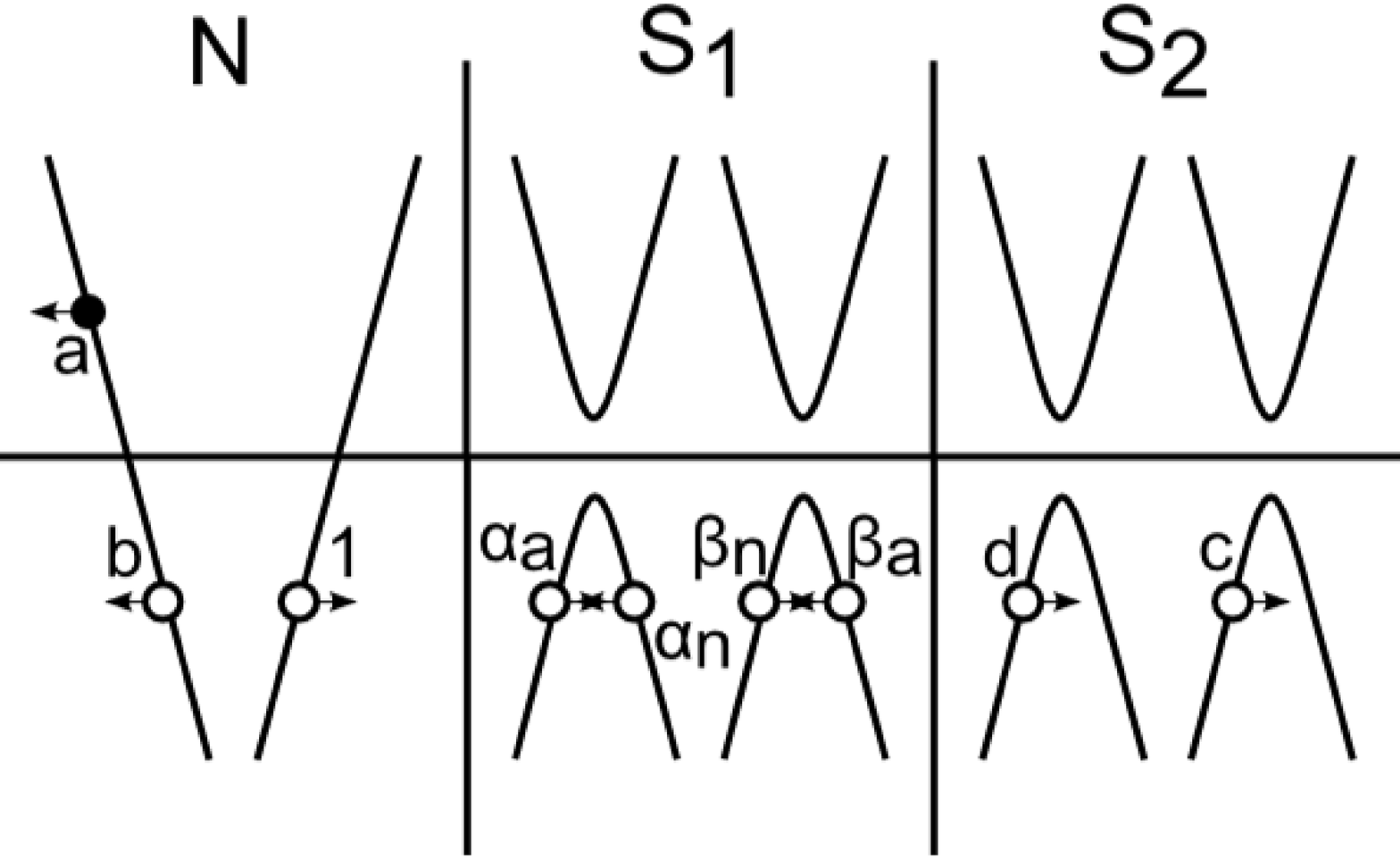}
\caption{\label{incident_h}Semiconductor diagram for an incident hole at the
  double {$NIS_1IS_2$} interface. The amplitudes $a,b,...$ are related to those of Fig. \ref{incident_e} by symmetry.}
\end{figure}

\begin{equation} \label{eq_psi_h}
\begin{split}
\Psi_N^{h} \left(x\right) & = \left( \begin{matrix} 0 \\ 1 \end{matrix}
\right) \left( e^{-i q^{-} x} + \bar b\, e^{iq^{-}x} \right)+\left( \begin{matrix}
  1 \\ 0 \end{matrix} \right) \bar a\, e^{-iq^{+}x} \\ \Psi_{S_1}^{h}
\left(x\right) & = \left( \begin{matrix} u_0 \\ v_0 \end{matrix} \right)
\left( \bar {\alpha}_n\, e^{i k^{-} x} + \bar {\beta}_n\, e^{-ik^{-}x} \right) \\ &+
\left( \begin{matrix} v_0 \\ u_0 \end{matrix} \right) \left( \bar {\alpha}_a\, e^{i
  k^{+} x} + \bar {\beta}_a\, e^{-ik^{+}x} \right) \\ \Psi_{S_2}^{h} \left(x\right) &
= \left( \begin{matrix} u_0 e^{i \varphi} \\ v_0 \end{matrix} \right) \bar c\,
e^{-i k^{-} x} + \left( \begin{matrix} v_0 e^{i \varphi} \\ u_0 \end{matrix}
\right) \bar d\, e^{ik^{+}x}.
\end{split}
\end{equation}

Denoting as $B^{ee}(E)=b(E)b^*(E)$ and $A^{eh}(E)=\bar a(E)\bar a^*(E)$  the
normally reflected probability and the Andreev reflected one (from a hole at energy $E$ to an electron at energy $-E$), the
quasiparticle current entering $S_1$ is given by\cite{BTK}

\begin{align}
\label{eq_current_asym}
I = 2 N(0) e \nu_F {\cal A} \int_{- \infty}^{\infty} &\left[ f_0 (E-eV) - f_0 (E) \right] \nonumber \\
 &\left[ 1+ A^{eh}(E) - B^{ee}(E) \right] dE,
\end{align}

\noindent
where $N(0)$ is the Fermi density of states in $N$, $\nu_F$ the Fermi velocity and ${\cal A}$ the junction area. $f_0 (E)$ is the equilibrium Fermi distribution. 
Thus the differential conductance is at zero temperature (${\cal N}=N(0)\nu_F {\cal A}$ is the number of channels):

\begin{equation}
G(E)=\dfrac{d I}{d V} = \dfrac{2 e^2{\cal N}}{h} \left[ 1 + A^{eh}(E) - B^{ee}(E) \right].\label{eq_conductance_asym}
\end{equation}

\begin{table*}
\begin{tabular}{c||c|c|c|c|}
{} & Figure No &$(E\rightarrow -E)$&$(\varphi\rightarrow
-\varphi)$&$(E,\varphi)\rightarrow(-E,-\varphi)$\\
\hline\hline
$Z_1$ or $Z_2=0$, $E<\Delta$ & \ref{Z_1=Z_2=0_L=1},\ref{Z1=0_or_Z2=0},\ref{addfig}a & yes& yes& yes\\
\hline
$Z_1$ or $Z_2=0$, $E>\Delta$ & \ref{Z_1=Z_2=0_L=1},\ref{Z1=0_or_Z2=0},\ref{addfig}a & no& no& yes\\
\hline\hline
$Z_1$ and $Z_2\ne0$, $E<\Delta$ & \ref{Z_1=Z_2=1_L=3},\ref{addfig}b & yes& yes& yes\\
\hline
$Z_1$ and $Z_2\ne0$, $E>\Delta$ & \ref{Z_1=Z_2=1_L=3},\ref{addfig}b & no& no& no\\
\hline\hline
\end{tabular}
\caption{
\label{table}
Summary of the different symmetries for the conductance.
}
\end{table*}

 The reference chemical potential is the one of $S_{1,2}$. The solutions {to Eqs. (\ref{eq_psi_e}) and (\ref{eq_psi_h})}
 are obtained by matching {the} wavefunction and its
 derivative at the interface in a standard
   procedure \cite{BTK}. Following the Andreev approximation, the wavevectors 
$k^+ \approx k^- \approx q^+ \approx q^- \approx k_F$ as factors in the derivatives 
of the wavefunctions, but their full expressions are kept in the exponentials. A complete analytical solution can be obtained, 
but it is too lengthy to be reported here. Yet it  can be used to check certain symmetry properties. For instance, the time-reversal symmetry 
is obeyed, manifesting here in the relation $A^{eh}(E, \varphi, k^{+,-})=A^{he}(-E,-\varphi,-k^{+,-})$ and $B^{ee}(E, \varphi, k^{+,-})=B^{ee}(E,-\varphi,-k^{+,-})$. 
The sign change in the phase reflects that of the (orbital) magnetic field, and time symmetry also inverts momenta 
as apparent from the matching equations (\ref{eq_psi_e}) and (\ref{eq_psi_h}).

\subsection{Analytical results in limiting cases}
An analytical solution can be written {for perfectly
  transparent contacts} ($Z_1=Z_2=0$). Such perfect contacts are an ideal limiting regime, that can be approached with quantum point contacts. 
For an incoming
electron, the solution reads
\begin{gather}
a = \dfrac{v}{u} \dfrac{u^2 \left( e^{i \varphi} - 1 \right) - \left( u^2 e^{i
    \varphi} - v^2 \right) e^{-i \kappa L}}{v^2 \left( e^{i \varphi} - 1
  \right) - \left( u^2 e^{i \varphi} - v^2 \right) e^{-i \kappa
    L}} \label{eq_a_e} ,\\ c = \dfrac{1}{u} \dfrac{\left(v^2 - u^2 \right)
  e^{-i \kappa L}}{v^2 \left( e^{i \varphi} - 1 \right) - \left( u^2 e^{i
    \varphi} - v^2 \right) e^{-i \kappa L}} \label{eq_c_e}
,
\end{gather}
and $b=d=0$, where $\kappa = k^{+} - k^{-} = \dfrac{k_F}{\mu} \sqrt{E^2 -
  \Delta^2}$.

 {For} energies {larger} than $\Delta$, $e^{-i
   \kappa L}$ is {in general} a complex number whereas for
 energies lower than $\Delta$, $e^{-i \kappa L}$ is {always a}
 real number . One verifies easily that
 $A(E)=1$ for $E < \Delta$, a result thus insensitive to $\varphi$
 (see Fig. \ref{Z_1=Z_2=0_L=1}).

The situation is different for $E>\Delta$. {The condition
  $\kappa L = 2n \pi$ expresses the occurrence} of constructive interferences between the two waves $k^{+}$ and
$k^{-}$ within the width of $S_1$.  This leads to the
  BTK-like scattering amplitudes $a= \dfrac{v}{u}e^{-i\varphi},c =
  \dfrac{1}{u}e^{-i\varphi},b=d=0$.

The condition $\kappa L = 2n \pi$ reads
\begin{equation}
E^2 = \Delta^2 + \dfrac{4 n^2 \pi^2 \mu^2}{L^2 k_F ^2},
\end{equation}

\noindent
which indicates the maxima in the Tomasch oscillations.

Another interesting limiting case is that of high energies $E
  \gg \Delta$, with $u^2 \simeq 1$ and $v^2 \simeq
0$. Then,
\begin{equation}
a \simeq \dfrac{v}{u} \left[(e^{-i \varphi} - 1)e^{i \kappa L} +1\right]
\end{equation}
thus
\begin{equation}
A \simeq   \left|  \dfrac{v}{u} \right|^2 \left[ 3 - 2 \cos \left( \varphi \right) - 2 \cos \left( \kappa L \right) + 2 \cos \left( \varphi - \kappa L \right) \right]
 \label{eq_max_phi}.
\end{equation}
{Eq.~(\ref{eq_max_phi}) leads to} $A \left(\varphi = \pi,
\kappa L \right)/A \left(\varphi = 0, \kappa
L \right)$ proportionnal to $5-4 \cos \left(\kappa L \right) \geq 1$ for all $\kappa L$. Thus, the conductance at
$\varphi = \pi$ is larger than the conductance at $\varphi = 0$
if $E\gg\Delta$.
(see Fig. \ref{Z_1=Z_2=0_L=1}). This trend helps to
understand the more general results presented later on.

\begin{figure}
\centering
\includegraphics[width=6cm]{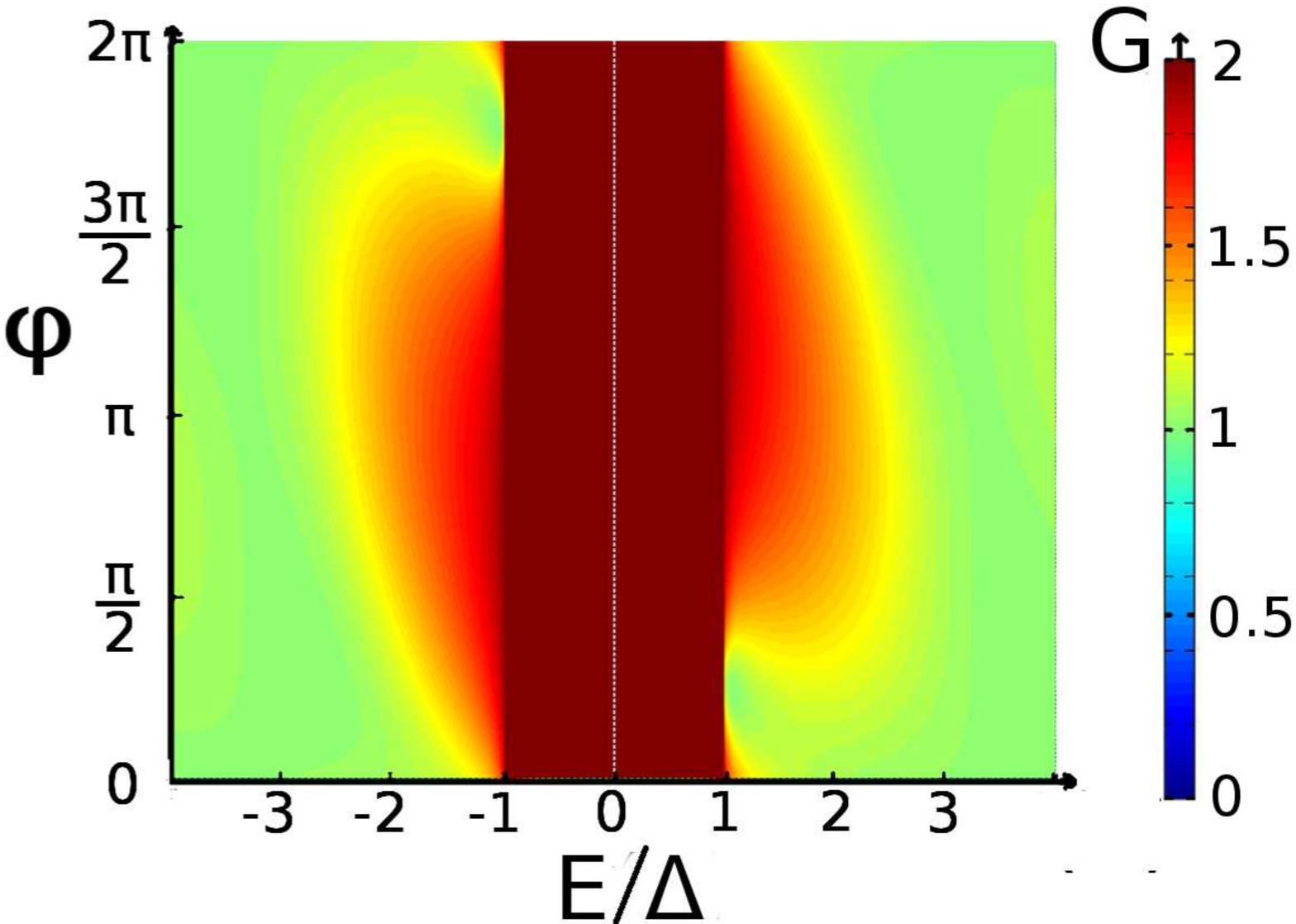}
\caption{\label{Z_1=Z_2=0_L=1}Map of the conductance $G(E,\varphi)$ (color on
  line). Case of fully transparent interfaces, $Z_1=Z_2=0$,
  $L/\xi=1$. $\Delta/\mu=0.002$ as well as in all other figures. No phase sensitivity is obtained below the gap.
  A phase-sensitive conductance enhancement is obtained above
    the gap, with a symmetry in the transformation $(E,\varphi \rightarrow -E,
    -\varphi)$.}
\end{figure}

On the other hand in the case of an incoming hole, one finds that
\begin{gather}
\bar a = \dfrac{u}{v} \dfrac{v^2 \left( e^{i \varphi} - 1 \right) - \left( u^2 e^{i
    \varphi} - v^2 \right) e^{-i \kappa L}}{u^2 \left( e^{i \varphi} - 1
  \right) - \left( u^2 e^{i \varphi} - v^2 \right) e^{-i \kappa
    L}} \label{eq_a_h}\\ 
\bar c = \dfrac{1}{v} \dfrac{\left(v^2 - u^2 \right) e^{-i
    \kappa L}}{u^2 \left( e^{i \varphi} - 1 \right) - \left( u^2 e^{i \varphi}
  - v^2 \right) e^{-i \kappa L}} \label{eq_c_h}
\end{gather}
and $\bar b=\bar d=0$. Then the question of changing sign of both $(E, \varphi)$
arises. Indeed if one changes $u \rightarrow v$, $v \rightarrow -u$
and $ \varphi \rightarrow - \varphi$ in Eqs.~\eqref{eq_a_e}
  and~\eqref{eq_c_e}, one finds that the moduli of $a$ and $c$ in
Eqs.~\eqref{eq_a_e}
  and~\eqref{eq_c_e}, and of $\bar a$ and $\bar c$ in
Eqs.~\eqref{eq_a_h}
  and~\eqref{eq_c_h} are respectively
equal. There is no symmetry under inversion of $E$ or
$\varphi$ separately, but there is symmetry under
simultaneous inversion of {$(E,\varphi)$}
(see Fig. \ref{Z_1=Z_2=0_L=1}). As shown below, this does
not hold any longer if $Z_1$ and $Z_2$ are both nonzero.

\begin{figure}
\centering
\begin{tabular}{cc}
\includegraphics[width=4cm]{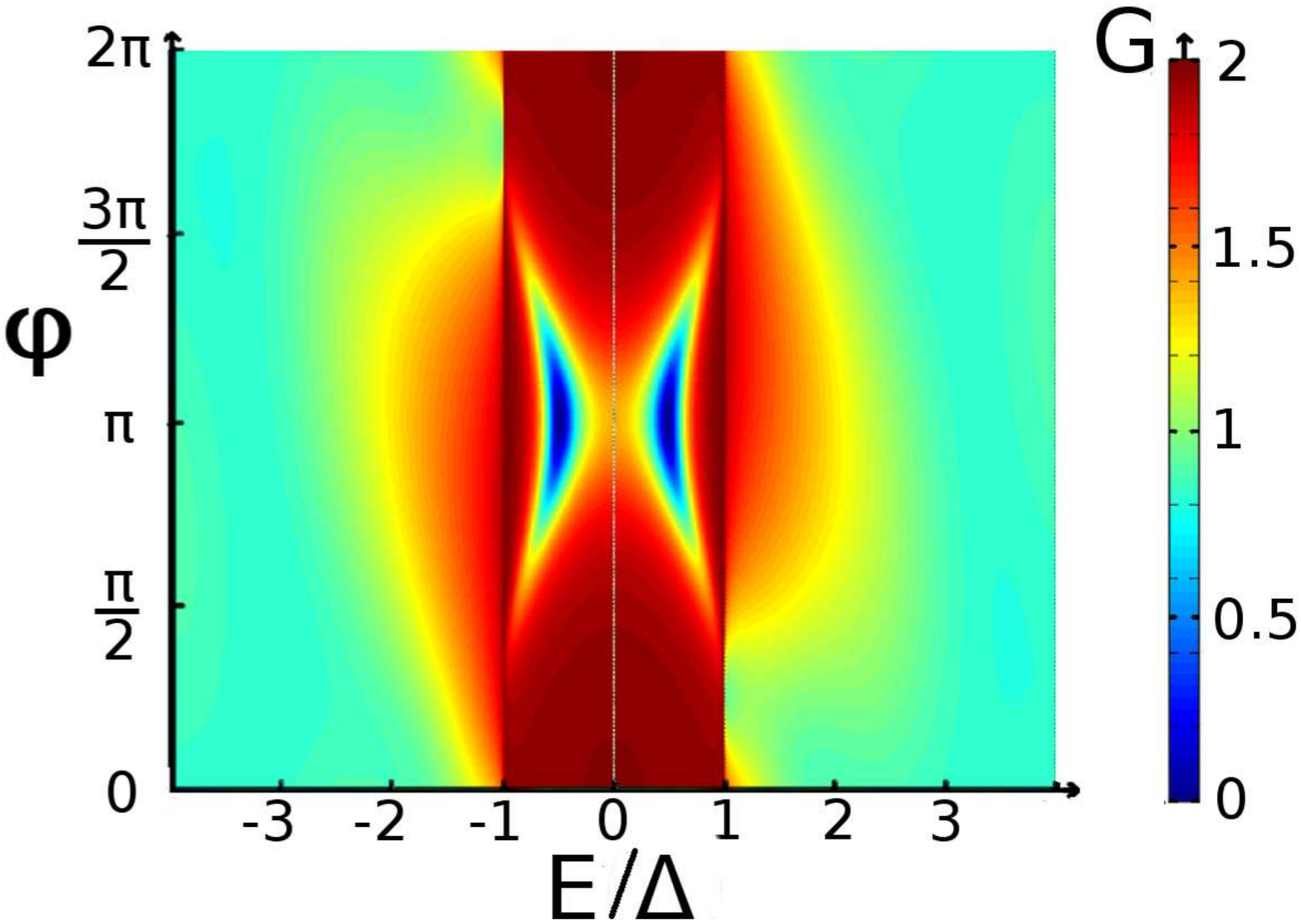}
&
\includegraphics[width=4cm]{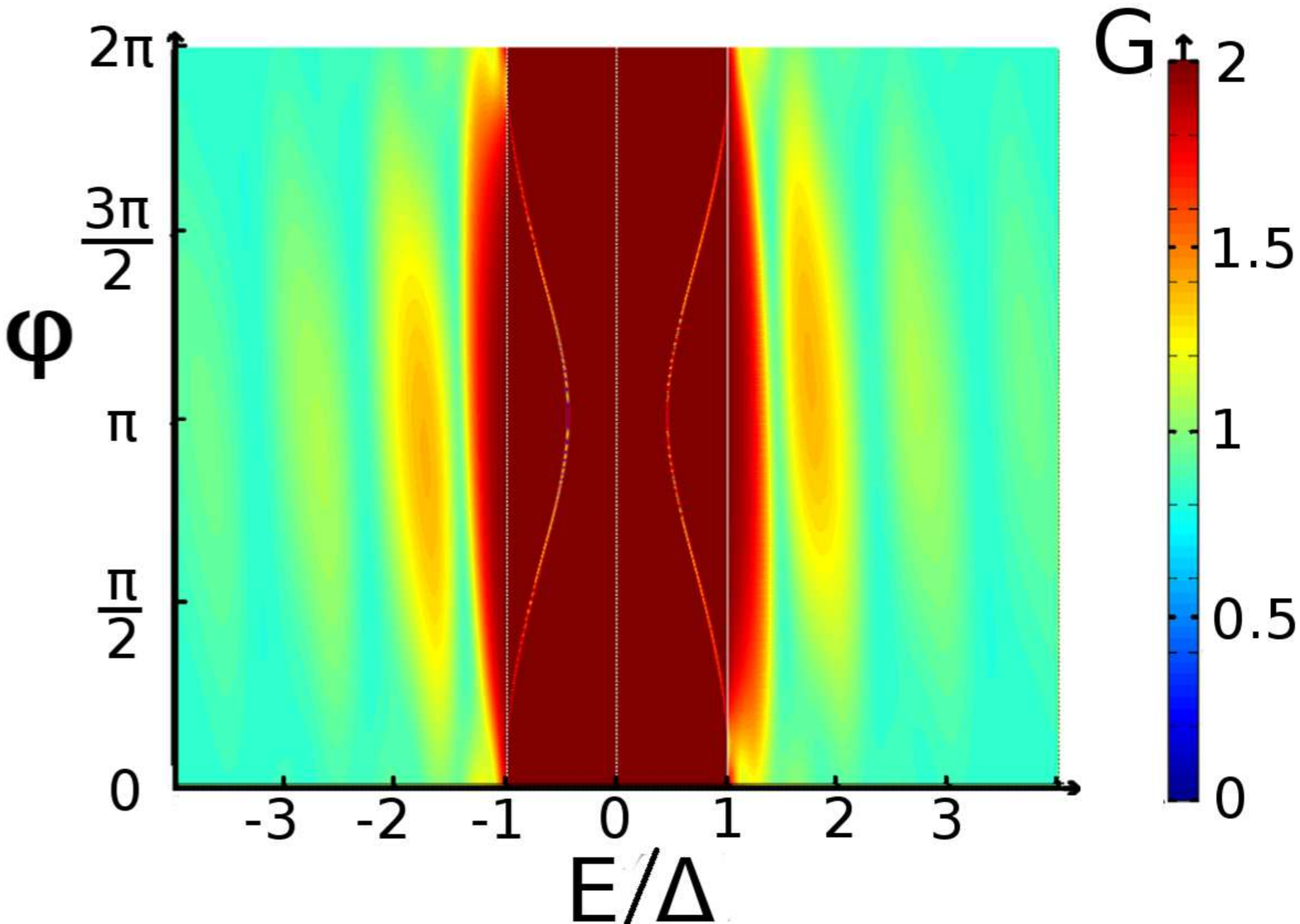}
\\
a) & b)
\\
\includegraphics[width=4cm]{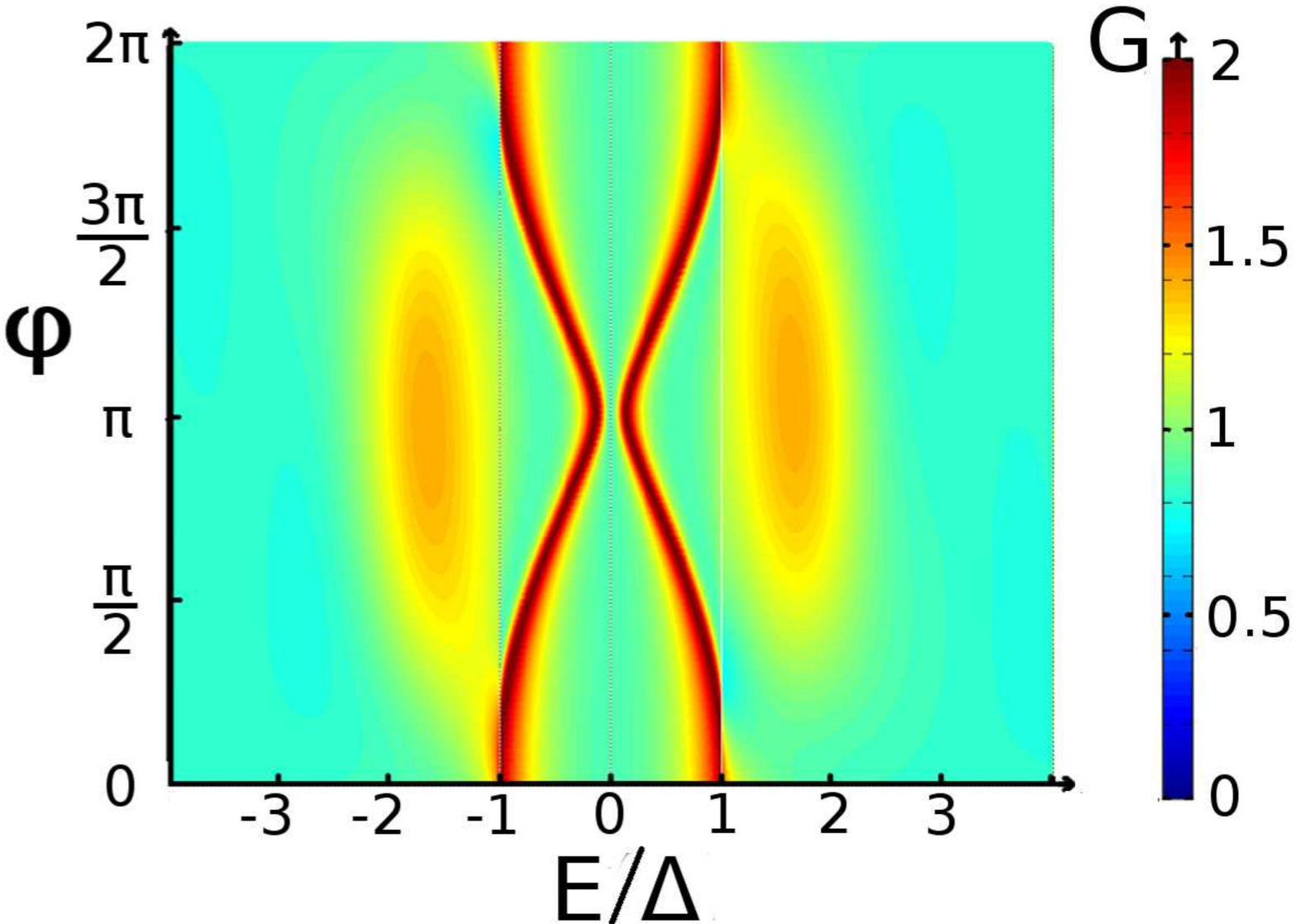}
&
\includegraphics[width=4cm]{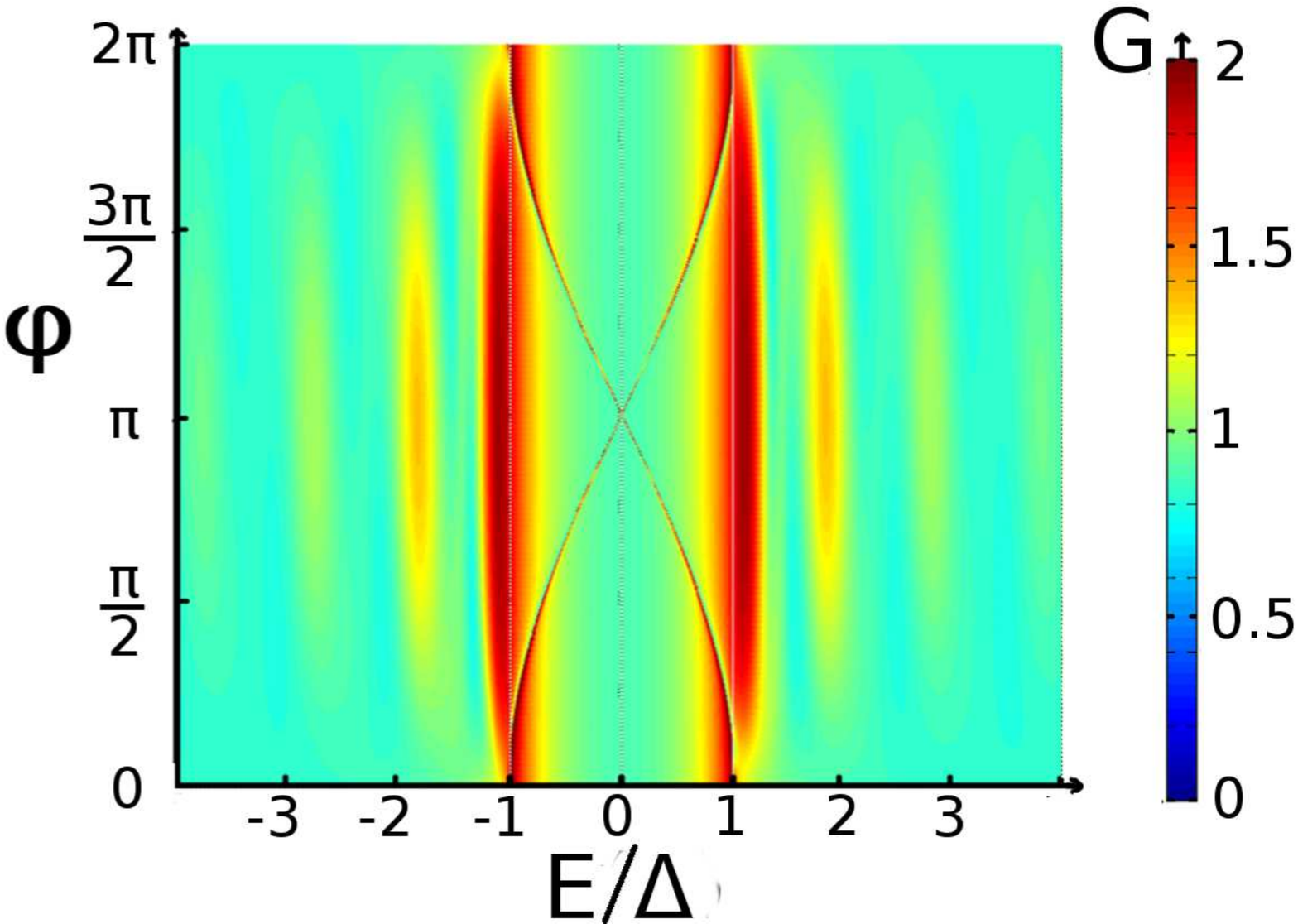}
\\
c) & d)
\\
\end{tabular}
\caption{\label{Z1=0_or_Z2=0}One interface only is transparent. Phase and
  energy symmetry is obtained below the gap, and phase-sensitive Tomasch
  oscillations above the gap, being symmetric in the
    transformation $(E,\varphi \rightarrow -E, -\varphi)$. a) $Z_1=0$,
  $Z_2=0.5$, {$L/\xi=1$}. b) $Z_1=0$, $Z_2=0.5$,
  {$L/\xi=3$}. {c) }$Z_1=0.5$, $Z_2=0$,
  {$L/\xi=1$}. {d) }$Z_1=0.5$, $Z_2=0$,
  {$L/\xi=3$}. In b), d),  the conductance
  anomaly at the gap edge is shifted by the phase towards
  $E>\Delta$, and the Tomasch oscillations are amplified.}
\end{figure}

\section{Conductance above the gap}
Let us first discuss how the excess conductance due to
Andreev reflections above the gap is modified by the phase difference
$\varphi$. As shown above for perfectly transparent
  interfaces, the conductance at $\varphi=\pi$ can be much larger than at
  $\varphi=0$. This behavior holds also for arbitrary $Z_1$, $Z_2$, as seen
  from the forthcoming discussion.

First, taking either ($Z_1=0$, $Z_2\neq0$) or ($Z_1\neq0$,
  $Z_2=0$), Fig.  \ref{Z1=0_or_Z2=0}
shows maps of the conductance as a function of energy
$E$ and phase $\varphi$ for several values
of $L/\xi$.   Symmetry of the conductance
  between ($E,\varphi$) and ($-E,-\varphi$) is obtained in each case.
 Second, the distinguishing features of Tomasch interferences
  appear as fringes in Fig.~\ref{Z1=0_or_Z2=0}. Their enhancement by a
  phase difference is visible and culminates at $\varphi=\pi$.
Moreover, the
  conductance just above the gap is markedly modified by the phase $\varphi$:
  the gap edge anomaly is shifted to higher energies, with a maximum for
  $\varphi=\pi$ , instead of a gap edge anomaly exactly at $E=\Delta$ for a
  single NS interface\cite{BTK}.

\begin{figure}[htb]
\centering
\begin{tabular}{cc}
\includegraphics[width=4cm]{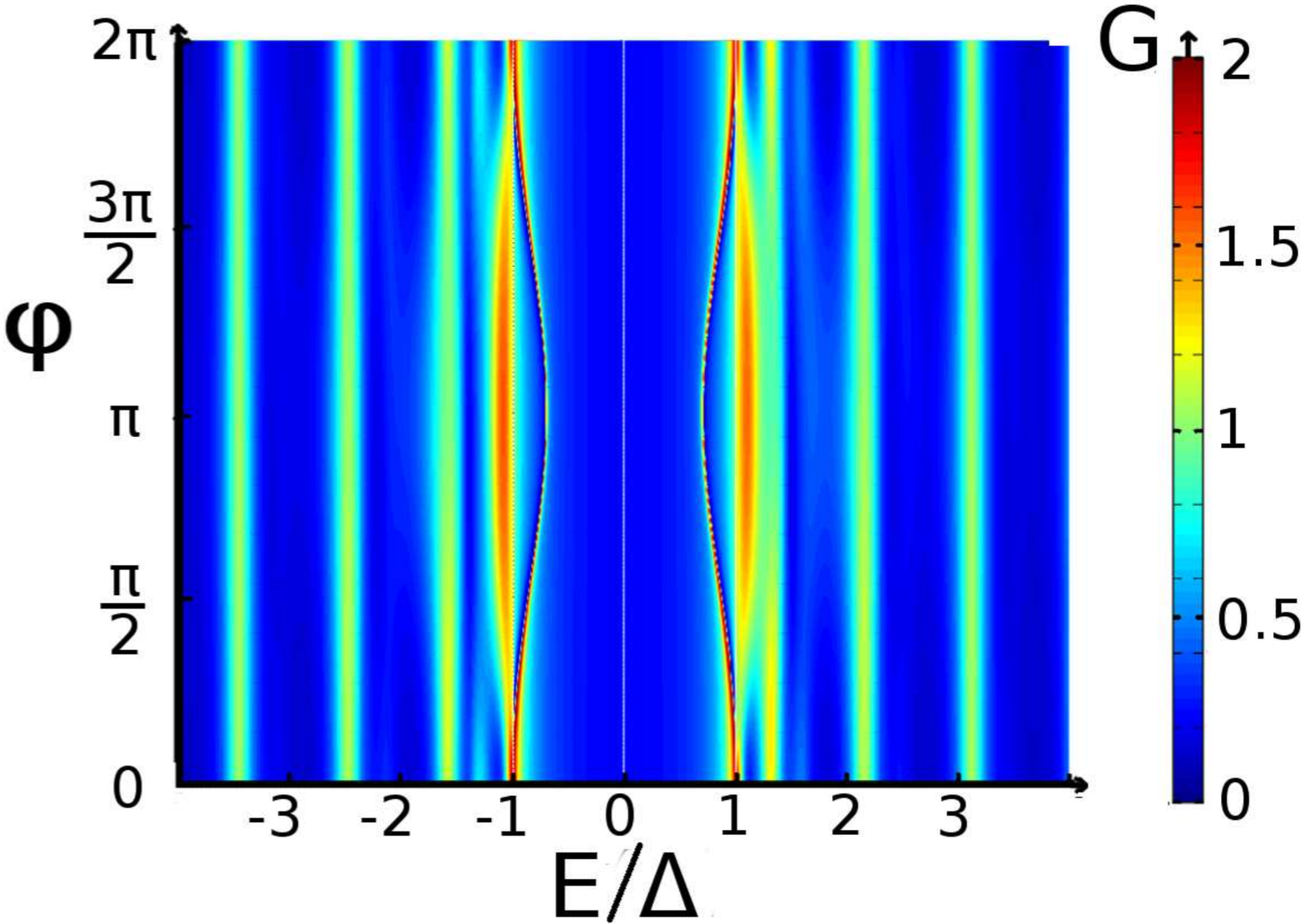}
&
\includegraphics[width=4cm]{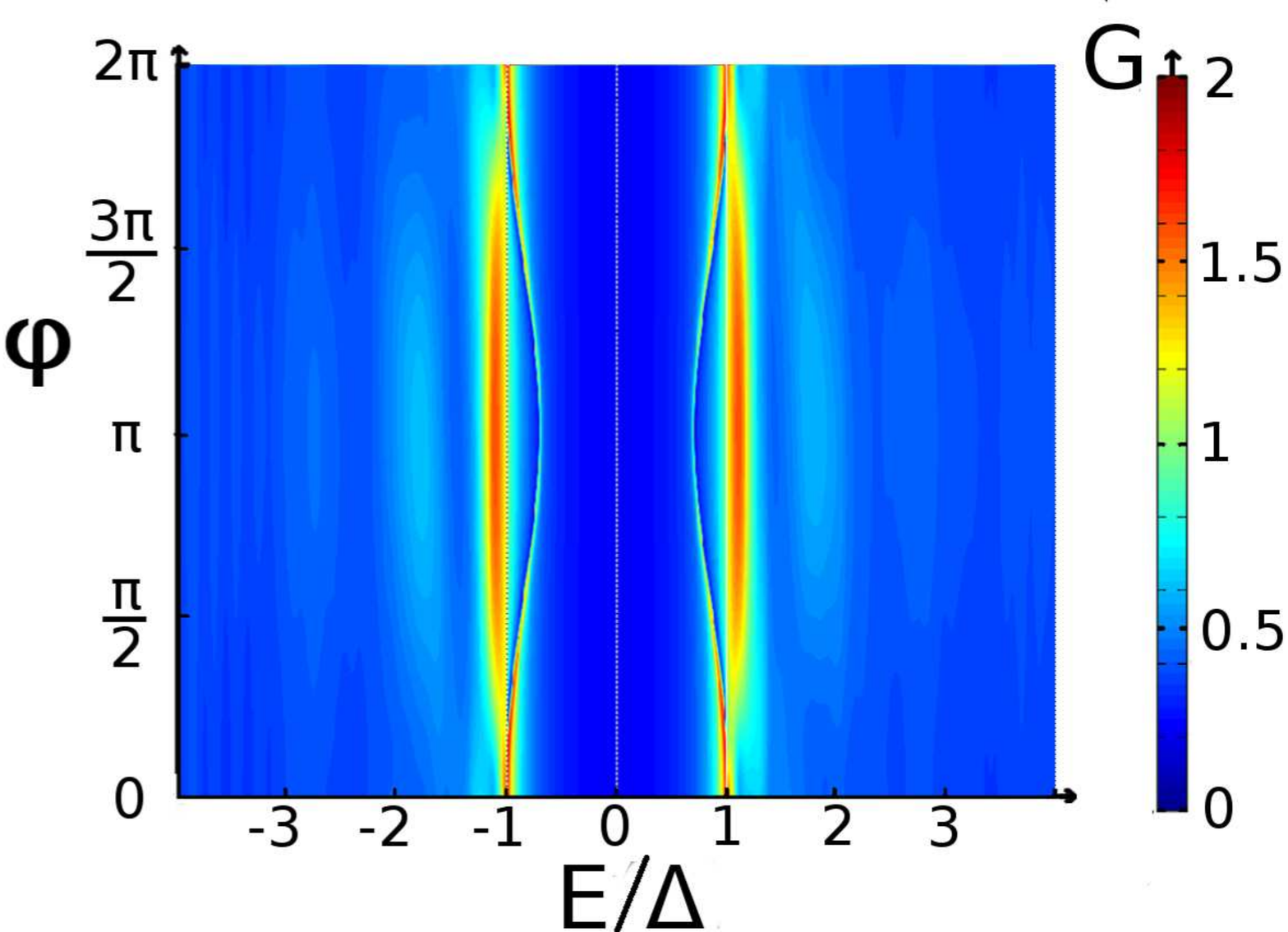}
\\
a) & b)
\\
\end{tabular}
\caption{\label{Z_1=Z_2=1_L=3}a) Symmetric barrier case, with $Z_1=Z_2=1$,
  $L/\xi=3$. b) Same parameters, but averaging $L$ on an
  interval comparable to $\xi$.}
\end{figure} 

If none of the barriers is perfectly transparent,
e.g. generically $Z_1\neq0$ and $Z_2\neq0,$ then no symmetry exists in
energy and phase ({see} Figs. \ref{Z_1=Z_2=1_L=3}a,
\ref{addfig}b). As mentioned above, this is not contradictory with the existence of a time inversion symmetry, but should be traced back to
the combination of several relevant phase shifts : those at the interfaces, related to $Z_1$ and $Z_2$, the wavevector phase shifts $(k^+L,k^-L)$ 
and the $\varphi$-dependent Andreev phase shifts at each of the two interfaces. 
Careful examination of the analytical solution of Eqs. (\ref{eq_psi_e}) and (\ref{eq_psi_h}) shows that the probabilities 
$A$ and $B$ are formed  by two kinds of terms. Those involving the phase shift $(k^+-k^-)L$, happen to be even in $Z_1$ and $Z_2$, and they are 
symmetric in $E, \varphi \rightarrow -E, -\varphi$. 
When $Z_1$ and $Z_2$ are both different, new terms containing 
$(k^{+}+k^{-})L\sim 2k_FL$ also appear, which are no more even in $Z_1$ and $Z_2$, and lead to breaking of the symmetry, e.g. 
$G(E,\varphi)\neq G(-E,-\varphi)$. This differences oscillates as $\cos 2k_FL$, and is expected to disappear with disorder, interface roughness 
or simply two-dimensional character of the interfaces. 

  On the other hand, when $Z_1$ and $Z_2$ are comparable, the
behavior for $|E|\gg\Delta$ features a Fabry-P\'erot-like cavity,
with quasi-periodic fringes extending at high energy (see
Fig. \ref{Z_1=Z_2=1_L=3}a). Since in a real experiment with extended
interfaces, the length $L$ is expected to fluctuate at the scale of the Fermi
wavelength $\lambda_F=\frac{2\pi}{k_F}$, those fringes are
expected to partially average out, as seen on Fig. \ref{Z_1=Z_2=1_L=3}b.

\section{Subgap conductance}

\subsection{Small {$L\alt\xi$} case}
The subgap conductance exhibits very interesting structures. First,
if $L \alt \xi$, and with a transparent $NS_1$ interface, most
of the subgap conductance is suppressed for phases close to $\pi$, as for
instance for $Z_1=0, Z_2=0.5$ (see Fig. \ref{addfig}a). This
is an interference phenomenon, culminating at $\varphi=\pi$ where the
amplitudes of Andreev reflections at $S_1$ and $S_2$ are just opposite to each
other. For small $L/\xi$, this destructive interference
holds in most of the subgap domain. For $L=\xi$ (see
Fig. \ref{Z1=0_or_Z2=0}a) or larger, it concentrates on a narrow energy
interval, a phenomenon that we discuss below.

\begin{figure}[htb]
\begin{tabular}{cc}
\includegraphics[width=4cm]{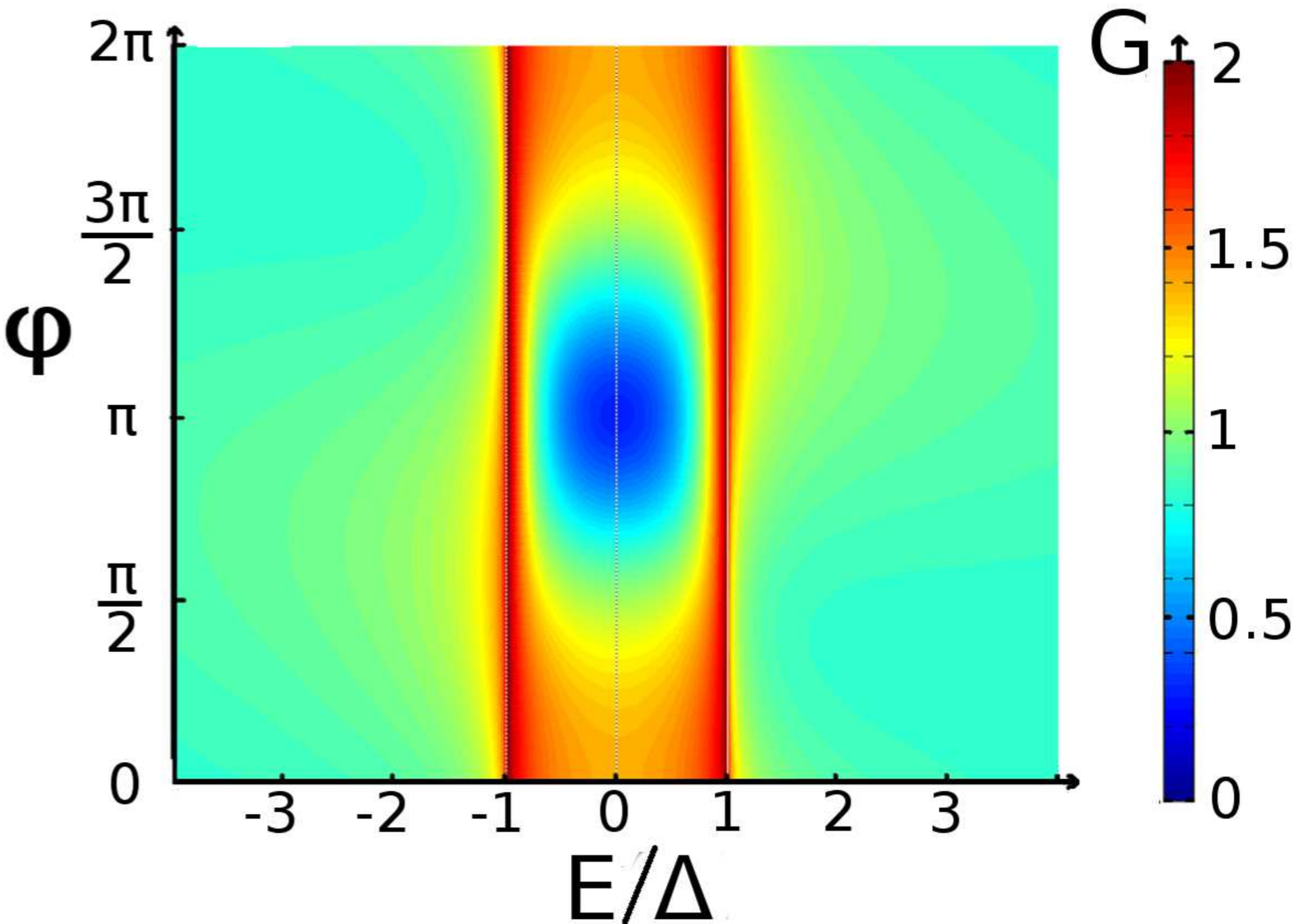}
&
\includegraphics[width=4cm]{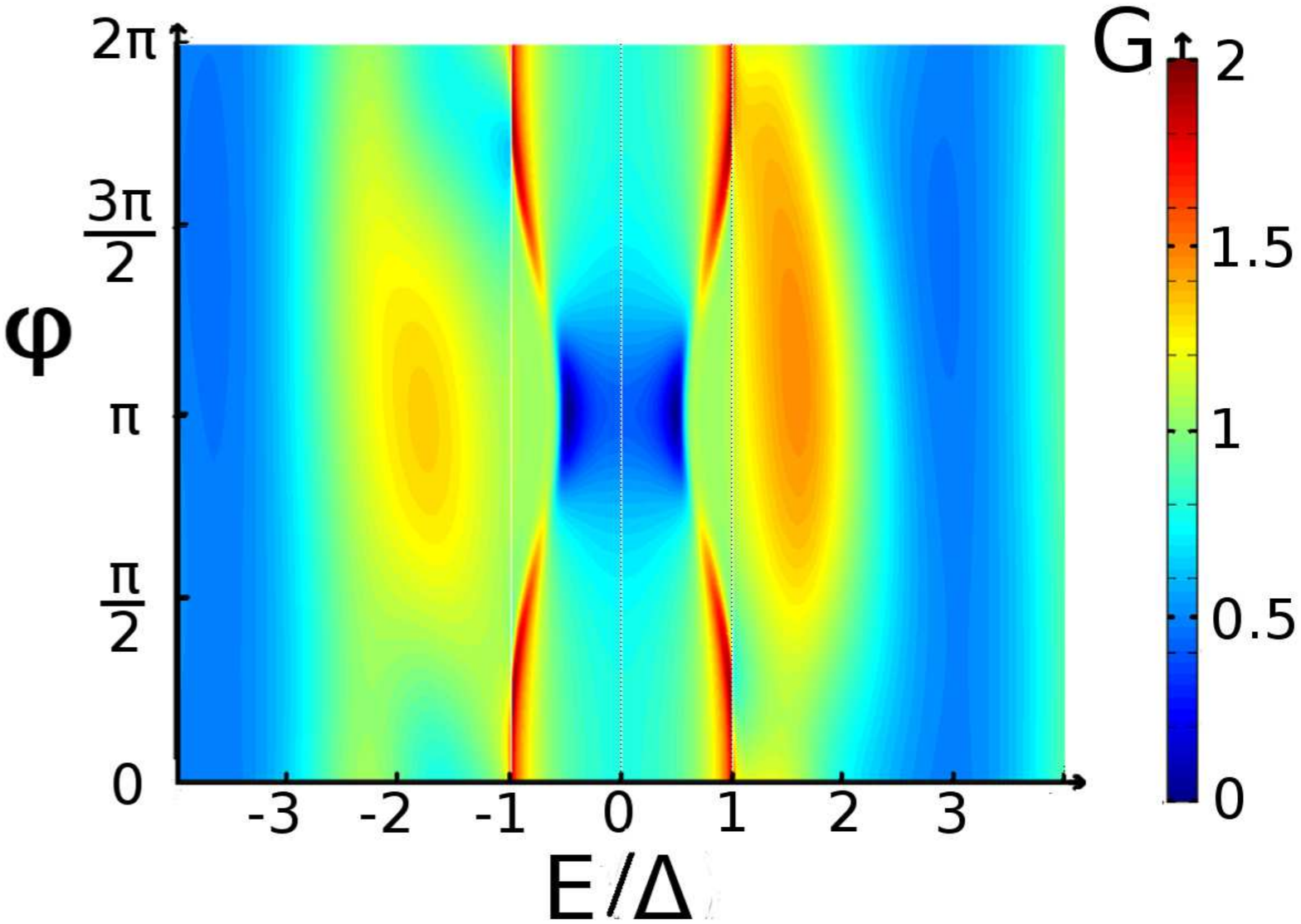}
\\
a) & b)
\\
\end{tabular}
\caption{\label{addfig}a) Asymmetric barrier case, with $Z_1=0$, $Z_2=0.5$,
  $L/\xi=0.2$. For a length $L$ shorter than $\xi$,
  interferences strongly decrease the subgap conductance{. b)}
  Both nonzero barriers, $Z_1=0.5$, $Z_2=0.5$,
  $L/\xi=1$, no symmetry in energy and phase is obeyed above
  the gap for a fixed $k_FL$.}
\end{figure} 

If on the other hand if the first interface is less
transparent, for instance $Z_1=0.5, Z_2=0$, the behavior for
  $L \alt \xi$ displays a pinching of the Andreev resonance anomaly (see
Fig. \ref{Z1=0_or_Z2=0}c), shifted at energies lower than the gap, with a
minimum at $\varphi=\pi$, going to zero energy if $L$ is
  large compared to $\xi$.

\subsection{Large $L\agt\xi$ case}

\subsubsection{Numerical results}
For small $Z_1$, together with the shift of the Andreev maximum towards above
the gap (see Fig. \ref{Z1=0_or_Z2=0}b), there appears a
conductance dip inside the subgap region with high conductance. Conversely, in
the opposite case where $Z_1>Z_2$ (see
Fig. \ref{Z1=0_or_Z2=0}d), the anomaly follows a similar energy and phase
variation but is dominated by a conductance excess.

\subsubsection{Analytical results for the conductance maxima and minima}

Those trends are better understood by plotting the subgap conductance as a
function of energy, for instance at $\varphi=\pi$ and
$L/\xi=3$. Fixing $Z_2=1$ and varying $Z_1$ from
$Z_1=0$ to $Z_1=2$ shows a drastic evolution
(see Fig. \ref{Conductance_Z2=1}). For $Z_1=0$, a sharp
conductance minimum appears, reaching zero. As $Z_1$ increases, a conductance
maximum develops in addition, and it dominates the anomaly for
$Z_1=2$. The conductance at the minimum is equal to zero
only if $\varphi=\pi$. For other values of the phase, the conductance minimum
does not reach zero.

\begin{figure}
\centering
\begin{tabular}{cc}
\includegraphics[width=4cm]{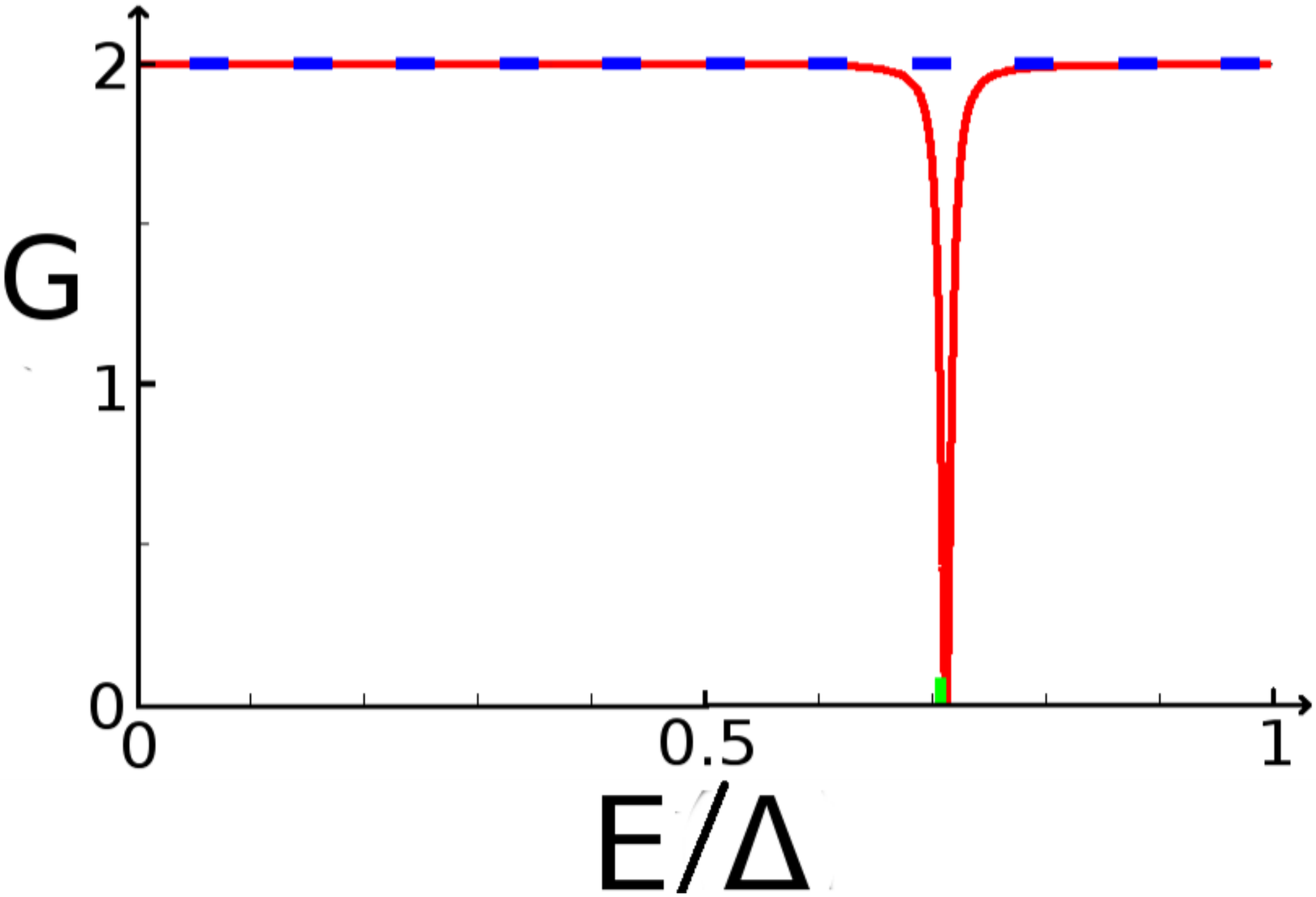}
&
\includegraphics[width=4cm]{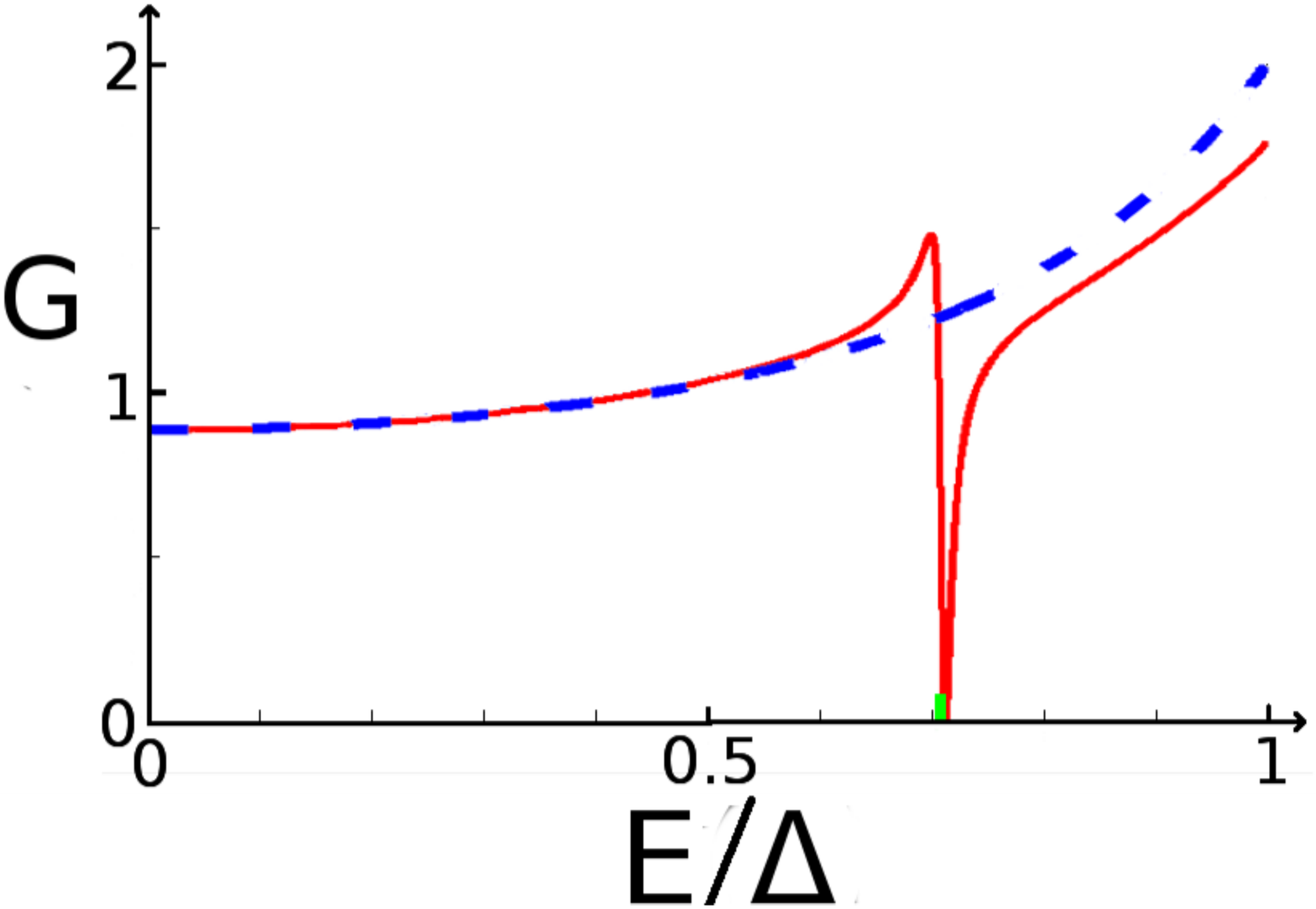}
\\
a) & b)
\\
\includegraphics[width=4cm]{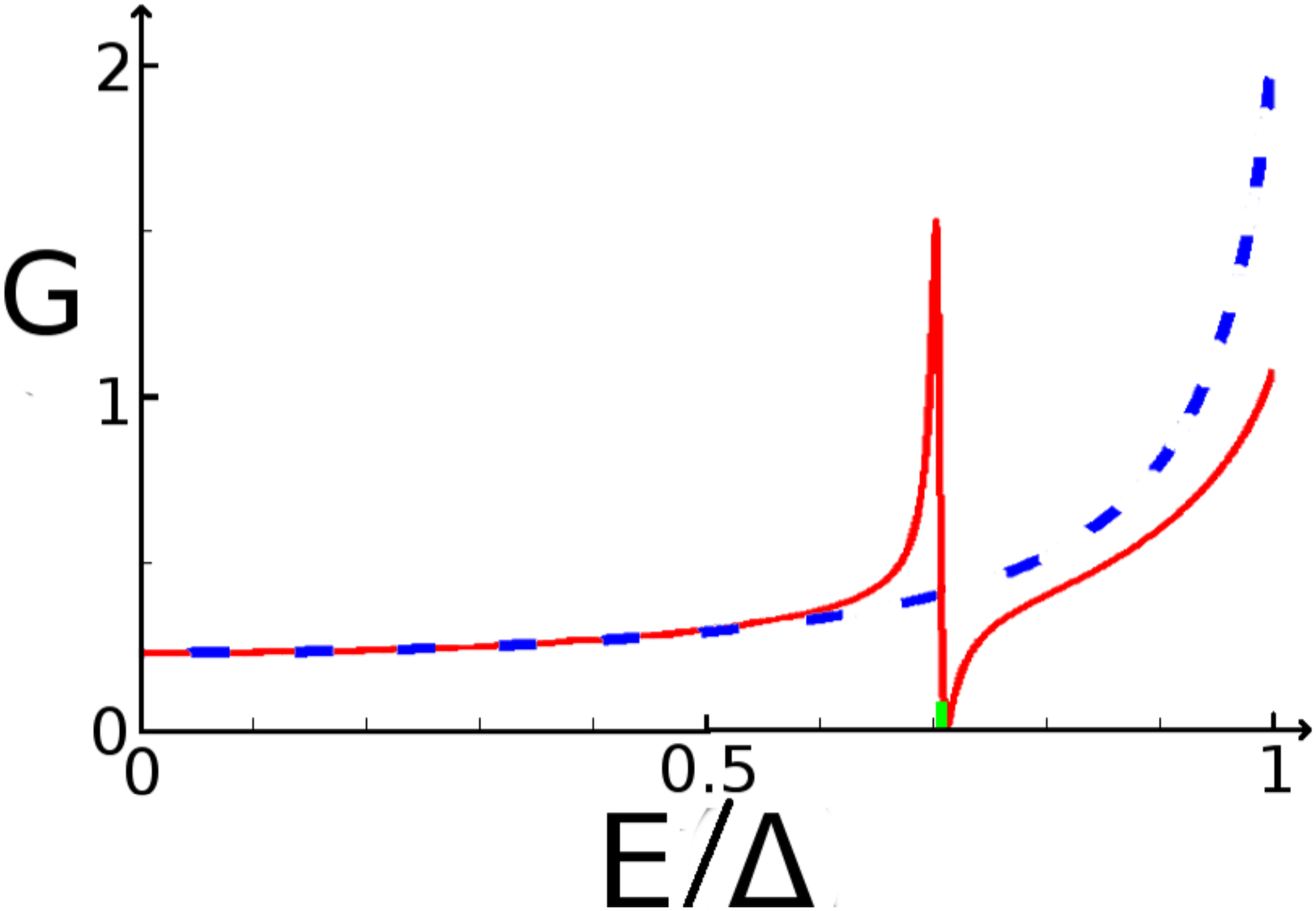}
&
\includegraphics[width=4cm]{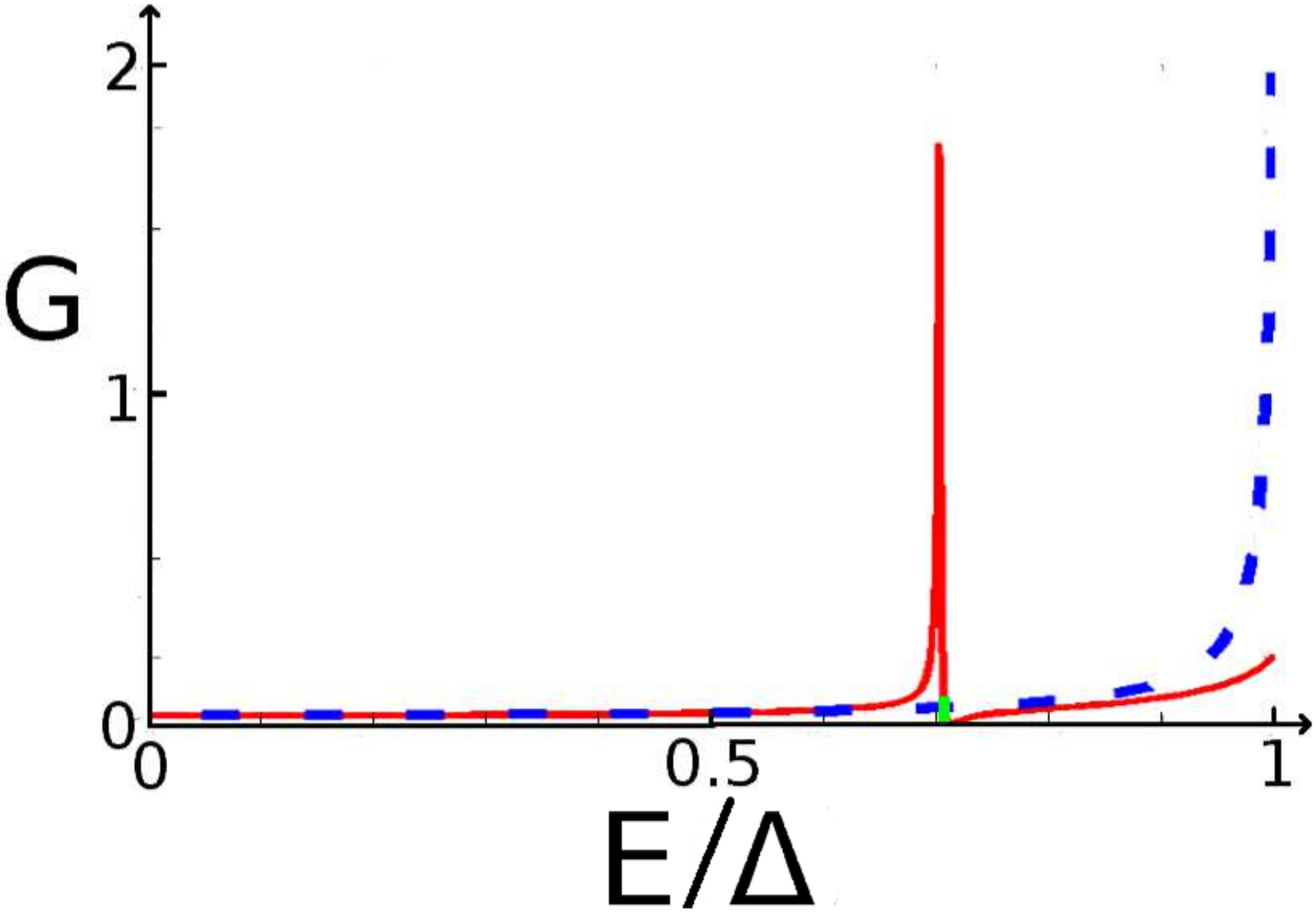}
\\
c) & d)
\\
\end{tabular}
\caption{\label{Conductance_Z2=1}Conductance curve for $Z_2=1$,
  {$L/\xi=3$}, $\varphi= \pi$ and different values of
  $Z_1$. Red (continuous line): the conductance of the NSS structure. Blue
  (dotted line): Conductance of the NS junction alone ($Z_1$ barrier). The zero of G signals the ABS energy for a barrier $Z_2$. a)
  $Z_1=0$ b) $Z_1=0.5$ c) $Z_1=1$
  {d)} $Z_1=2$.}
\end{figure}

An analytical insight of both the minimum and maximum of the conductance can
be obtained. Let us first set $Z_2=0$. Then one looks for zeros of the normal
reflection coefficient $b$, meaning that Andreev reflection is total at the
$NS_1$ interface, in spite of $Z_1 \neq0$. One gets an energy-phase condition
(here $x={v_0^2}/{u_0^2}$):

\begin{widetext}
 \begin{equation} \label{eq_phi(E)_general}
 \varphi = - i \ln \left( \dfrac{e^{4\kappa L} \left(x^2 +1\right) -2x \pm \sqrt{e^{4 \kappa L} \left(x^2 -1 \right) \left( \left(x^2 +1 \right) e^{4 \kappa L} -4x \right)}}{2x \left( e^{4 \kappa L} -1 \right)} \right).
 \end{equation} 
 \end{widetext}
 
If $L$ is large compare to $\xi$,
Eq.\eqref{eq_phi(E)_general} simplifies into $\varphi = \mp i \ln \left( e^{-2i \arccos \left(
  E/\Delta \right)} \right)$, and one obtains
\begin{equation}
E_{ABS} = \pm \Delta \cos\left( \dfrac{\varphi}{2} \right)
,
\end{equation}
corresponding to the ABS energy for a transparent single
channel $S_1S_2$ contact (or many degenerate channels for a planar interface). If the interface
between $N$ and $S_1$ is not perfectly transparent and
if $L$ is large compared to $\xi$, the normal metal
{electrode} $N$ can act like a side
tunneling probe, with a conductance maximum
  at the the energy of the ABS localized at the $S_1S_2$
interface (see Figs. \ref{Z1=0_or_Z2=0}d and
\ref{Conductance_Z2=1}d). Otherwise, if $L \lesssim \xi,$ the barrier at the
 $NS_1$ interface perturbs more strongly     
the reflections involved in the ABS, and an energy gap opens
between negative and positive-energy ABS, as observed in Fig
\ref{Z1=0_or_Z2=0}c, together with substantial broadening.

An unexpected behavior is encountered
if the barrier $NS_1$ is {\it more} transparent than the Josephson
barrier $S_1S_2$. As shown in Figs \ref{Conductance_Z2=1}a, b), the
spectroscopic signature of the ABS is a conductance {\it
  minimum}. Again, an analytical solution can be obtained for $Z_1=0$, looking
for maxima (instead of zeros) in the reflection coefficient $b(E)$. This yields
\begin{equation}
e^{i \varphi } \left( Z_2^2 + 1 \right) \left(x^2+1\right) - x \left(2 e^{i
  \varphi} \left( Z_2^2 + 1 \right) + \left( 1 - e^{i \varphi} \right) ^2
\right) = 0
\end{equation}
where $x = u_0^2/v_0^2$. 
Solving this equation, one finds
\begin{equation}
\label{eq:A}
E_{ABS} = \pm \Delta \sqrt{1 - T \sin^2 (\varphi / 2)}
\end{equation}
where {$T = 1/(1+Z_2^2)$} is the Josephson junction
transparency. { Eq.~(\ref{eq:A})} is the {ABS
  energy} for a single channel with barrier $Z_2$.

\section{Discussion and conclusion}
\subsection{Discussion of the results}

The above results show that Andreev scattering at a $NS_1S_2$ interface
displays a rich behavior if the width of $S_1$ is
comparable to the coherence length and a phase difference can be applied at
the junction $S_1S_2$. Above the gap, the interferences between quasiparticle
modes propagating within $S_1$ become phase-sensitive, which
enhances the Tomasch oscillations with a maximum 
at $\varphi=\pi$.

An interference occurs between the Andreev scattering amplitudes at the two
interfaces for subgap voltage. At large
$L/\xi$, a sharp resonance appears at the energy of the
ABS. To be observable, the transparency of the Josephson junction should be large enough, 
so that the ABS extends inside the superconducting gap. The structure of this resonance displays a maximum and
a minimum of conductance, separated by a small energy
  difference (see
Fig. \ref{Conductance_Z2=1}). The relative
weight of the maximum and minimum depends
  on the respective transparencies of the two interfaces. When $Z_1>Z_2$,
it is dominated by an enhanced transmission, in a way similar to tunnel
spectroscopy. The normal reflection amplitude has a minimum and the Andreev
reflection amplitude a maximum.  Conversely, when $Z_1<Z_2$, a destructive
interference occurs in the Andreev channel. For $\varphi=\pi$, this
interference completely cancels Andreev reflection, thus the conductance
becomes zero, and this is even true for any value of $Z_1$, $Z_2$ and $L$,
except for very small values $L \ll \xi$ (see
Fig. \ref{Conductance_Z2=1}). This spectacular result means that the
superconductor becomes opaque to quasiparticles coming from the normal
metal. Symmetrically, quasiparticles involved in ABS at the
$S_1S_2$ Josephson junction are completely reflected at the $NS_1$
interface. For phases different from $\pi$, reflection is partial, but a sharp
minimum occurs and this is enough to detect the ABS.  Let us stress that the larger $L$, the sharper the resonance,
which exists even for $L\gg\xi$. This is is due to the
divergence of the effective scattering length at the resonance, in a way
similar to the usual Andreev resonance at the gap edge in the case of a single
NS interface.

It is possible to interpret the conductance maximum and
  minimum by diagrams in the corresponding limiting cases.   
Fig. \ref{diagramme}a show a constructive interference
between Andreev reflections at $NS_1$ and $S_1S_2$, where multiple Andreev
reflections builds the ABS. On the other hand,
Fig. \ref{diagramme}b shows a constructive interference
between normal reflections at $NS_1$ and multiple Andreev reflections at
$S_1S_2$. The perfect reflection at $NS_1$ implies the formation of a resonant
state within the superconductor, which involves a Cooper pair crossing
$S_1S_2$ and a crossed Andreev reflection ($e-h$ line in
Fig. \ref{trion}). This three-body process can also be viewed as an exchange
process between a single quasiparticle and one member of a Cooper pair,
accompanied by a pair crossing the junction.

\begin{figure}
\begin{tabular}{cc}
\includegraphics[width=4cm]{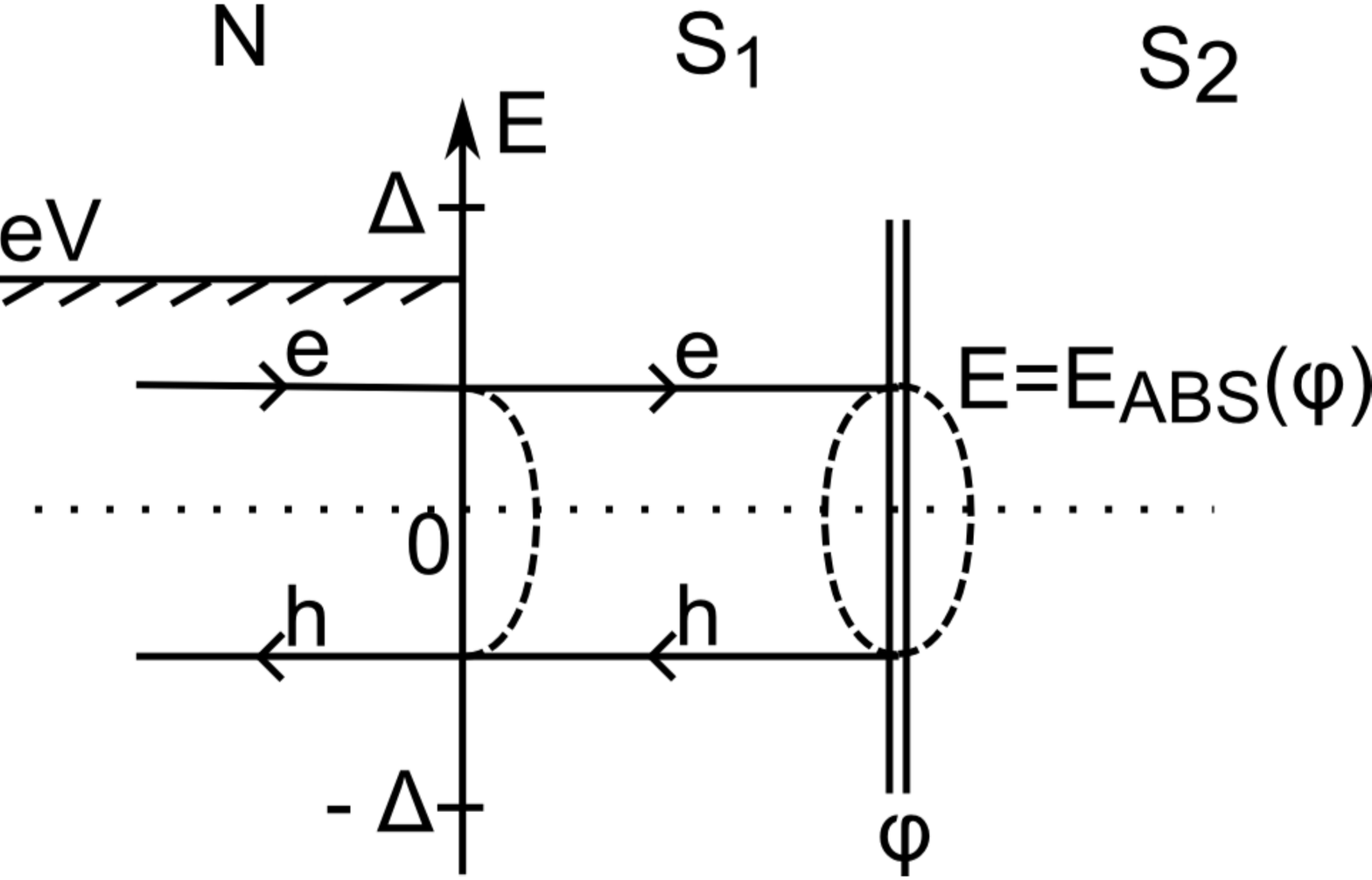}
&
\includegraphics[width=4cm]{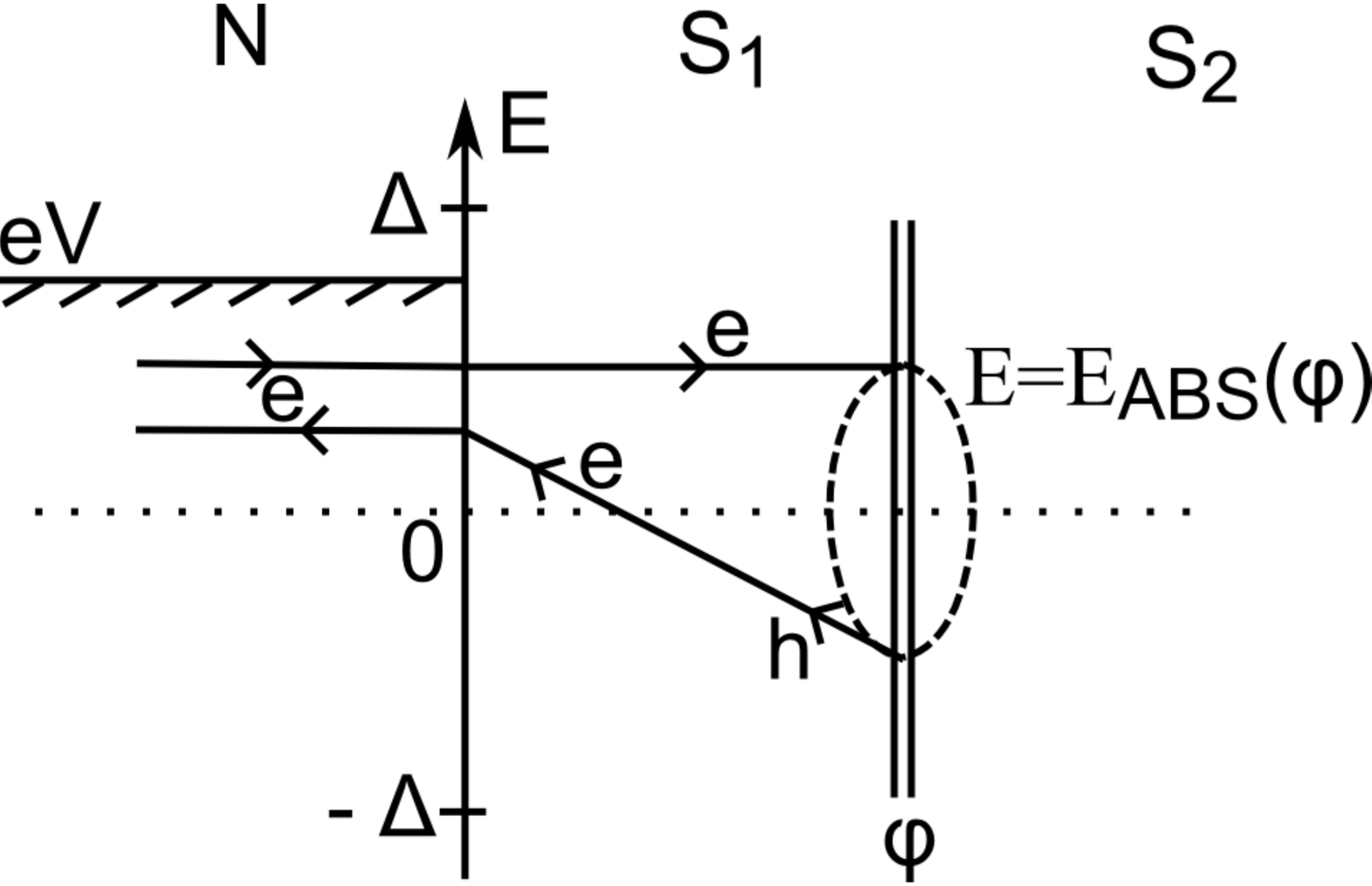}
\\
a) & b)
\\
\end{tabular}
\caption{\label{diagramme} a) Scattering diagram showing the cooperative
  Andreev reflections at the $NS_1$ interface and at the Josephson $S_1S_2$
  junction, dominating in the large $Z_1$ case, and responsible for the conductance maximum
  (Fig. \ref{Conductance_Z2=1}c,d). b) Scattering diagram showing the normal reflection resulting of the
  combination of Andreev scattering at $S_1S_2$ and crossed Andreev process, 
dominating in the small $Z_1$ case, and
  responsible for the conductance minimum of Figs. \ref{Conductance_Z2=1}a,b.}
\end{figure}

\begin{figure}
\includegraphics[width=8cm]{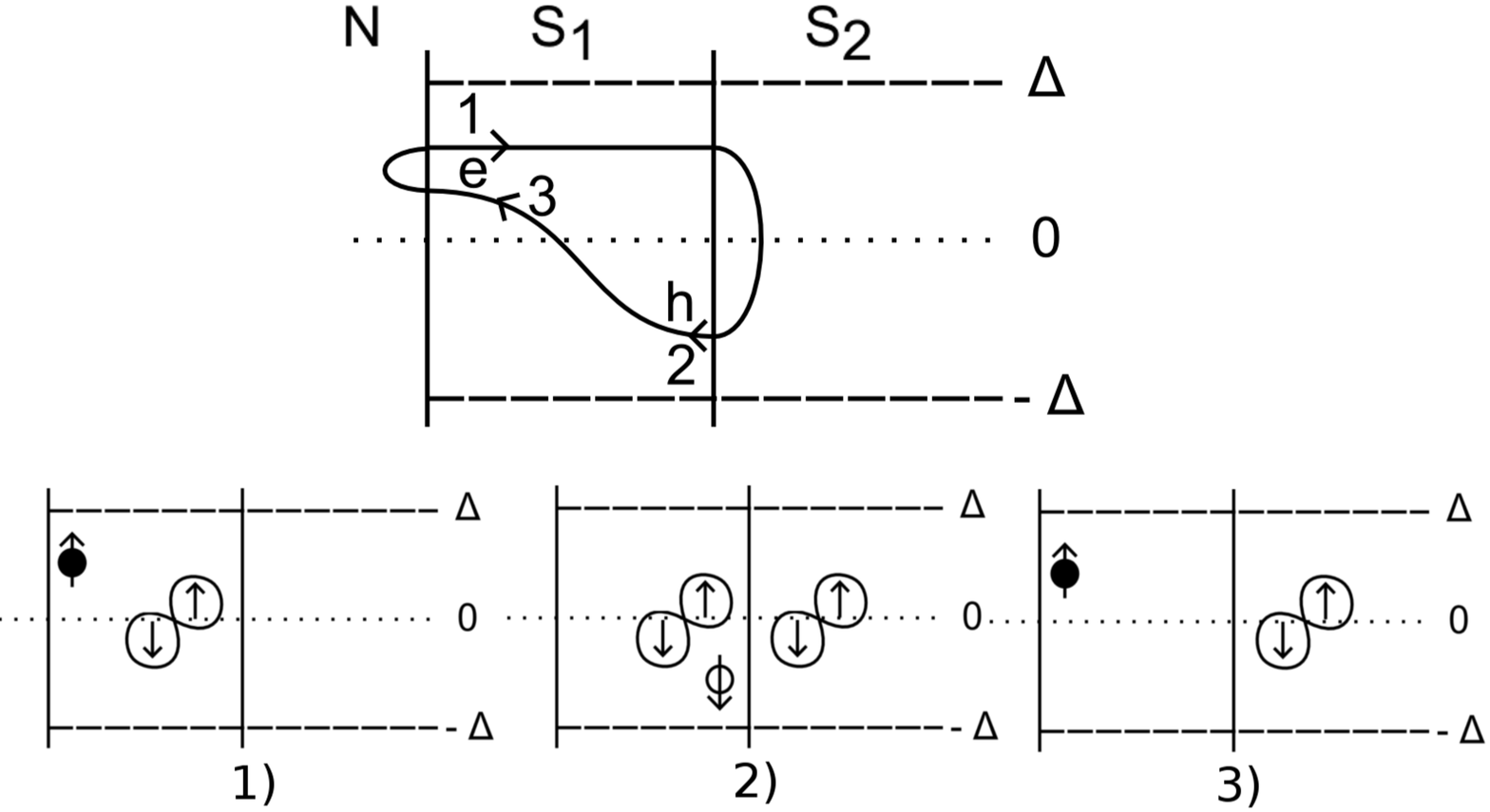}
\caption{\label{trion} Diagram showing the resonant state formed with a Cooper
  pair crossing the $S_1S_2$ junction and a quasiparticle trapped in $S_1$ (small 
$Z_1$ (Fig. \ref{Conductance_Z2=1}a,b). Steps 1-3
  achieve exchange between a single quasiparticle and a Cooper pair crossing
  the junction : First, an electron (in black) at the $NS_1$ interface
  propagates and is reflected into a hole at $S_1S_2$, while a Cooper pair is
  transferred at the junction; Second, the reflected hole propagates back to
  $NS_1$ into the original electron, in a crossed Andreev process.}
\end{figure}

The situation {$Z_1\ll Z_2$}, where
the spectroscopic signature of the Andreev states is a conductance minimum, is
especially interesting. Then the quasiparticle current flowing at the $NS_1$
interface can be larger than the critical current at the $S_1S_2$
junction. The scattering approach does not ensure conservation of the
quasiparticle current, since quasiparticles are converted into Cooper
pairs. The excess current in $NS_1$ compared to $S_1S_2$ should flow in the upper branch of the set-up (Fig. \ref{scheme}). 

\subsection{Multichannel or two-dimensional contacts, and effects of disorder}
The above analysis considers the most simplified one-dimensional model, at zero temperature. It can describe parallel interfaces with 
the same number of (nondispersive) channels. Or in the spirit of BTK \cite{BTK}, 
it can be taken as a phenomenological approach to a few channel point contact probed
by a tunneling tip in its close vicinity (or a Sharvin contact). Yet, the general trends {of the
  considered model} should {be revealed} in a more realistic set-up. First, at nonzero temperature, 
Fermi broadening of the electronic distribution in $N$ will smear the conductance anomalies reported in this work. Second, a multichannel
generalization of the scattering method is possible. If the $S_1S_2$ junction has many dispersive or diffusive channels, 
it defines ABS extending in energy above 
some minigap $\delta(\varphi)$, and one expects that the anomaly of the conductance will 
reveal the phase dependence of this minigap (this also holds if the junction 
is a diffusive $SNS$ junction). Depending on temperature and ABS level spacing, peaks or dips in the conductance can be resolved, 
or on the contrary merge into a shoulder (or trough) extending between $\delta(\varphi)$ and the gap $\Delta$. 

Disorder such as point disorder in $S$ or interface roughness is expected to have very different effects, depending on whether the voltage is larger or smaller than the gap. 
In the former case, disorder in the superconductor can easily blur the Tomasch oscillations, unless the elastic mean-free path is larger than $L$ (clean superconductor). 
In addition, it is expected to 
restore the symmetry $G(E,\varphi)=G(-E,-\varphi)$. In the subgap regime, on the contrary, the spectroscopic signatures of the ABS are pinned to the ABS energy and should 
be quite robust as suggested by averaging out fluctuations in $k_FL$ in the present calculation. Last but not least, disorder in the normal metal $N$ can amplify the Andreev reflection at low energy and give rise to subgap anomalies, by "reflectionless tunneling". To treat all these effects and perform more realistic
calculations, in terms of geometry and disorder, one requires more advanced methods
using nonequilibrium Green's functions. Such methods also allow to calculate the dependence with $V$ of the Josephson current, not addressed in this work. 
In the case of large transparency and many channels at the $NS_1$ interface, 
one should include self-consistency of the gap, and also possible
nonequilibrium effects.

In the case of the three-terminal geometry with different superconductors $S_1$ and $S_2$, those might have different gaps $\Delta_1 < \Delta_2$. 
One expects that the structure of the ABS below 
the smallest gap is revealed in $G(E,\varphi)$, and that a more complex behaviour is obtained between $\Delta_1$ and $\Delta_2$.

On the other hand, on the more classic tunnel spectroscopy case {$Z_1 \gg
  Z_2$}, where the probe little perturbs the junction, the present
configuration is advantageous in terms of spectroscopy of the ABS. This work suggests a
``side-spectroscopy'' by letting a scanning tunneling tip or narrow contact
come at a distance of order $\xi$ from the junction ({see}
Fig. \ref{scheme}).

In conclusion, we have revealed the rich behavior a double $NSS$ interface,
when the independent control parameters are a voltage bias and a
superconducting phase difference, respectively applied to the two
interfaces. Phase-sensitive Tomasch oscillations, together with various
spectroscopic probes of the {ABS}, are predictions that could
be tested in a realistic device.

The authors acknowledge support from the Agence Nationale de la Recherche
(contract Nanoquartets 12 BS10 007 01). The authors also thank H. Courtois, 
C. Winkelmann and B. Dou\c{c}ot for fruitful discussions.

\end{document}